\documentclass[a4paper,11pt]{article}
\pdfoutput=1
\usepackage{amsmath,amssymb,theorem,subcaption,tikz}
\usepackage{textgreek,mathrsfs,mathtools}
\usepackage[cm]{fullpage}
\usepackage[T1]{fontenc}
\usepackage[utf8]{inputenc}
\usepackage[shortcuts]{extdash}
\usepackage[unicode]{hyperref}
\usepackage[nosort]{cite}

\let\mathcal=\mathscr

\usetikzlibrary{shapes,decorations.pathmorphing}

\let\frac\undefined

\allowdisplaybreaks

\numberwithin{equation}{section}

\def\eq#1$$#2$${\begin{equation#1}#2\end{equation#1}}
\long\def\subeq#1{\begin{subequations}#1\end{subequations}}
\def\Split$$#1$${\begin{split}#1\end{split}}
\def\Align#1$$#2$${\begin{align#1}#2\end{align#1}}
\def\AlignAt#1$$#2$${\begin{alignat}{#1}#2\end{alignat}}
\def\Aligned#1{\begin{aligned}{}#1\end{aligned}}

\def\Gather#1$$#2$${\begin{gather#1}#2\end{gather#1}}
\def\Gathered#1{\begin{gathered}{}#1\end{gathered}}

\def\Multline#1$$#2$${\begin{multline#1}#2\end{multline#1}}

\def\Cases#1{\begin{cases}#1\end{cases}}
\def\d{\partial}
\def\bd{\bar\partial}

\def\Re{\mathop{\rm Re}\nolimits}

\def\sign{\mathop{\rm sign}\nolimits}

\def\const{\mathop{\rm const}\nolimits}

\def\cA{{\mathcal A}}

\def\cG{{\mathcal G}}

\def\cI{{\mathcal I}}

\def\cO{{\mathcal O}}

\def\cV{{\mathcal V}}

\def\hK{{\hat K}}
\def\hM{{\hat M}}
\def\hU{{\hat U}}
\def\hV{{\hat V}}
\def\ve{\varepsilon}
\def\sh{\mathop{\rm sh}\nolimits}
\def\ch{\mathop{\rm ch}\nolimits}

\def\tg{\mathop{\rm tg}\nolimits}
\def\ctg{\mathop{\rm ctg}\nolimits}

\def\Arg{\mathop{\rm Arg}\nolimits}
\def\tr{\mathop{\rm tr}\nolimits}
\def\lcolon{\mathopen{\,:}}
\def\rcolon{\mathclose{:\,}}
\def\C{{\mathbb C}}

\def\Z{{\mathbb Z}}
\def\cD{{\mathcal D}}
\def\P{{\mathbf P}}
\def\Q{{\mathbf Q}}
\def\a{{\mathbf a}}
\def\ba{\bar{\mathbf a}}

\def\T{\mathop{\rm T}\nolimits}

\def\e{{\rm e}}
\def\i{{\rm i}}

\def\bz{{\bar z}}

\def\bc{{\bar c}}
\def\bh{{\bar h}}

\def\reg{\text{reg}}

\def\thalf{{\textstyle{1\over2}}}
\def\lrd#1{\mathop{\mathord{\mathop\d\limits^{\lower.5ex\hbox{$\scriptstyle\leftrightarrow$}}}_{#1}}}

\def\txpsi{\text{\textpsi}}
\def\upstrut{\vrule height \ht\strutbox depth 0pt width 0pt\relax}

\makeatletter

\def\section{\@startsection{section}{1}{\z@}%
                                   {-3.5ex \@plus -1ex \@minus -.2ex}%
                                   {2.3ex \@plus.2ex}%
                                   {\normalfont\normalsize\bfseries}}
\def\subsection{\@startsection{subsection}{2}{\z@}%
                                     {-3.25ex\@plus -1ex \@minus -.2ex}%
                                     {1.5ex \@plus .2ex}%
                                     {\normalfont\normalsize\bfseries\itshape}}
\def\@seccntformat#1{\csname the#1\endcsname.~~}
\long\def\@makecaption#1#2{%
  \vskip\abovecaptionskip
  \sbox\@tempboxa{\small#1. #2}%
  \ifdim \wd\@tempboxa >0.9\hsize
  {\leftskip=0.05\hsize\rightskip=0.05\hsize\relax\small
    #1. #2\par}
  \else
    \global \@minipagefalse
    \hb@xt@\hsize{\hfil\box\@tempboxa\hfil}%
  \fi
  \vskip\belowcaptionskip}
\def\Appendix{\appendix
  \def\@seccntformat##1{Appendix~\csname the##1\endcsname.~~}}

\let\over\@@over
\let\atop\@@atop
\let\above\@@above
\let\overwithdelims\@@overwithdelims
\let\atopwithdelims\@@atopwithdelims
\let\abovewithdelims\@@abovewithdelims

\makeatother

\newtheorem{theorem}{Theorem}

\newtheorem{statement}{Statement}

\numberwithin{theorem}{section}
\numberwithin{lemma}{section}

\long\def\?#1{{\par\medskip\hrule\smallskip\noindent
{\bf What is missing:} #1\smallskip\hrule\medskip\par}}

\begin{document}
\title{Semiclassical approach to form factors in the sinh\-/Gordon model}
\author{Michael Lashkevich$^{a,b}$, Oleg Lisovyy$^c$, and Tatiana Ushakova$^d$\\
	\parbox[t]{.8\textwidth}{\normalsize\it\raggedright
	\begin{itemize}\itemsep=\smallskipamount
		\item[$^a$]Landau Institute for Theoretical Physics, 142432 Chernogolovka, Russia
		\item[$^b$]Kharkevich Institute for Information Transmission Problems, 19 Bolshoy Karetny per., 127994 Moscow, Russia
		\item[$^c$]Institut Denis\--Poisson, Université de Tours, CNRS, Parc de Grandmont, 37200 Tours, France
		\item[$^d$]Department of Physics and Astronomy, Johns Hopkins University, Baltimore, Maryland 21218, USA
	\end{itemize}
		ML:~lashkevi@landau.ac.ru, OL:~lisovyi@lmpt.univ-tours.fr, TU:~tushako1@jhu.edu}
}
\date{}

\maketitle

\begin{flushright}\it
Dedicated to the memory\\ of our friend and colleague\\ Yaroslav Pugai
\end{flushright}
\bigskip

\begin{abstract}
Form factors in the sinh\-/Gordon model are studied semiclassically for small values of the parameter $b\sim\hbar^{1/2}$ in the background of a radial classical solution, which describes a heavy exponential operator placed at the origin. For this purpose we use a generalization of the radial quantization scheme, well known for a massless boson field. We introduce and study new special functions which generalize the Bessel functions and have a nice interpretation in the Tracy\--Widom theory of the Fredholm determinant solutions of the classical sinh\-/Gordon model. Form factors of the exponential operators in the leading order are completely determined by the classical solutions, while form factors of the descendant operators contain quantum corrections even in this approximation. The construction of descendant operators in two chiralities requires renormalizations similar to those encountered in the conformal perturbation theory.
\end{abstract}

\section{Introduction}

Form factors in quantum field theory are matrix elements of local and quasilocal operators in the basis of eigenstates of the Hamiltonian of the theory. Studies of exact form factors in two\-/dimensional massive integrable models started at the end of seventies~\cite{Karowski:1978vz,Berg:1978sw}. F.~Smirnov~\cite{Smirnov:1984sx,Smirnov:1992vz} formulated a closed system of bootstrap equations for form factors for an integrable model with a given exact spectrum and a given exact $S$\-/matrix. Any solution to these bootstrap equations uniquely defines a (quasi)local operator. It was conjectured that form factors of any (quasi)local operator satisfy the bootstrap equations. The main difficulty of this approach is to identify local operators defined as solutions to the bootstrap equations with those defined in terms of the fields that enter the Lagrangian. This identification demands some additional reasoning, and had not been done completely for any theory except free field theories.

Form factors of local operators in the sinh\-/Gordon theory were studied extensively~\cite{Fring:1992pt,Mussardo:1992vg,Koubek:1993ke,Lukyanov:1997bp,Babujian:1999ht,Babujian:2002fi,Feigin:2008hs,Lashkevich:2013mca,Lashkevich:2013yja}. In fact, form factors were identified for exponential operators, energy\-/momentum tensor and, up to normalization and total curl terms, for conserved currents. Besides, for any operator with given form factors it is straightforward to obtain form factors of its commutator with any integral of motion. An important progress was related with the so\-/called fermion bases in the sine/sinh\-/Gordon theory~\cite{Jimbo:2011gv,Jimbo:2011bc}. This construction makes it possible to find the correspondence between solutions to the bootstrap equations and conformal descendant operators of the Liouville theory on at least first ten (and presumably on all) levels in each chirality through the mediation of the six\-/vertex lattice model~\cite{Boos:2009fs,Boos:2010ww,Negro:2013rxa,Boos:2016yql}. Despite the obvious success of this construction, it defines operators in the massless and massive case via different limits of lattice objects and does not clarify the relation between them in the framework of the field theory.

A massive model of quantum field theory can often be considered as a relevant perturbation of a conformal field theory~\cite{Zamolodchikov:1987jf,Zamolodchikov:1989zs}. Correlation functions of local operators in such model are described by the underlying conformal model at small distances, and thus have a power law decay in this region. At large distances, correlation functions decay exponentially, and their large distance asymptotics define the form factors of local operators. What happens with local operators under such perturbation was understood by Al.~Zamolodchikov~\cite{Zamolodchikov:1990bk} in the framework of the conformal perturbation theory. Formal correlation functions for many operators contain ultraviolet divergences. For the correlation functions to be finite, each of such operators should be renormalized by adding operators of sufficiently low conformal dimensions that appear in multiple operator product expansions of the original conformal operator with the perturbation operator. However, the conformal perturbation theory is not efficient at distances larger than the inverse mass, and thus cannot be used to extract form factors from correlation functions. The semiclassical approach, which we develop in the present paper, makes it possible to describe both small and large distances at the cost of the limitation to small values of the perturbation operator conformal dimension.

The sinh\-/Gordon theory can be thought either as a perturbation of the free massless boson, or as a perturbation of the Liouville theory. The Liouville theory possesses the so\-/called reflection property~\cite{Zamolodchikov:1995aa}. The reflection property identifies up to a factor the pairs of exponential operators $\e^{\alpha\varphi}$ and $\e^{(Q-\alpha)\varphi}$ with some value $Q$, which depends on the parameter of the theory. Since the sinh\-/Gordon model can be represented as a perturbation of the Liouville theory in two different ways, there are two reflection properties~\cite{Fateev:1997nn}: $\e^{\alpha\varphi}\sim\e^{(Q-\alpha)\varphi}\sim\e^{-(Q+\alpha)\varphi}$. We will see how the reflection properties appear in the semiclassical consideration.

The paper is organized as follows. In Section \ref{sec:radial-quant}, we describe the semiclassical limit for correlation functions that contain one exponential operator $\e^{\alpha\varphi(0)}$ and several field operators $\varphi(x_i)$. For such correlation functions, the field can be split into the sum of a classical radial solution and quantum fluctuations about it. It is convenient to describe the quantum part in terms of a radial quantization scheme, which generalizes the well\-/known radial quantization of the free massless boson field. The quantum field is expanded in a basis of functions that generalize the Bessel functions. In Section \ref{sec:fredholm-dets}, we use the Tracy\--Widom approach~\cite{Tracy:1996CMP179} to define these functions in a more universal way as derivatives of a Fredholm determinant solution to the classical sinh\-/Gordon equation with respect to auxiliary parameters. This makes it possible to describe their asymptotics and calculate certain integrals necessary to calculate vertices of Feynman diagrams. Then in Section~\ref{sec:form-factors} we arrive to our main goal and calculate a few series of form factors of local operators, including the exponential operators and some of the descendants. We construct the descendant operators by means of a rather standard regularization procedure on the Euclidean coordinate space: the derivatives of the field $\varphi$ are moved apart from the exponential field to a small distance. We see that in the case of chiral descendants this distance can be sent to zero, while the descendants that contain both right and left chiralities require a renormalization before taking this limit. All calculations are performed in the leading order, but the form factors contain both classical and quantum contributions.

\section{Radial quantization of the sinh\-/Gordon model in the semiclassical limit}
\label{sec:radial-quant}

Consider the quantum sinh\-/Gordon model with the Euclidean action
\eq$$
S_b[\varphi]={1\over8\pi}\int d^2x\,\left({(\d_\mu\varphi)^2\over2}+{m_0^2\over b^2}(\ch b\varphi-1)\right).
\label{shG-action}
$$
The model possesses a unique neutral particle of mass $m$ related to the parameter $m_0$ as follows~\cite{Zamolodchikov:1995xk}
\eq$$
m_0^2=16{\Gamma(1-b^2)\over\Gamma(1+b^2)}\left(m{\Gamma\left(1+{p\over2}\right)\Gamma\left(1-p\over2\right)\over4\sqrt\pi}\right)^{2+2b^2},
\qquad
p={b^2\over1+b^2}.
\label{m0-m-rel}
$$
Since the theory is integrable the $S$ matrix is purely elastic and factorizable, so that the scattering is fully characterized by the two\-/particle $S$ matrix
\eq$$
S(\theta)=-{\sin\pi p+\i\sh\theta\over\sin\pi p-\i\sh\theta},
\label{S-matrix}
$$
where $\theta=\theta_1-\theta_2$ is the difference of the rapidities of the colliding particles, i.e.\ the energy and momentum of the $i$th particle are given by $p^0_i=m\ch\theta_i$, $p^1_i=m\sh\theta_i$.

We will only consider the limit $b\ll1$, so that
\eq$$
m_0\simeq m,
\qquad
S(\theta)\simeq-\e^{\i\pi\sign\theta}
\label{m0-S-b-small}
$$
There are two remarks concerning these formulas. First, since all practical calculations in the present paper are performed in the leading order of $b$, we ignore the difference between $m$ and $m_0$, and write $m$ everywhere. It must be taken into account when applying our formulas to subleading contributions to correlation functions and form factors.

Second, formally the $S$ matrix tends to unity, so that the particle should become a free massive boson. Nevertheless, we have to remember that for any small but finite $b$ two particles cannot possess the same momentum due to the fact that $S(0)=-1$.

Define the operator
\eq$$
V_\nu(x)=\cG_\nu^{-1}\e^{b^{-1}\nu\varphi(x)},
\qquad
\cG_\nu=\langle\e^{b^{-1}\nu\varphi(0)}\rangle.
\label{Vnu-def}
$$
We will be interested in the correlation functions of the form
\eq$$
G_\nu({\{x_i\}_N})=\langle\varphi(x_N)\ldots\varphi(x_1)V_\nu(0)\rangle
=Z_\nu^{-1}\int\cD\varphi\,\varphi(x_N)\cdots\varphi(x_1)\e^{-S_b[\varphi]+b^{-1}\nu\varphi(0)},
\label{Gnu-def}
$$
where we used the shorthand notation $\{x_i\}_N$ for the ordered set $x_1,\ldots,x_N$.
Here
\eq$$
Z_\nu=\int\cD\varphi\,\e^{-S_b[\varphi]+b^{-1}\nu\varphi(0)}.
\label{Z-def}
$$
Consider the semiclassical limit $b\ll1$ of (\ref{Gnu-def}) keeping the value of~$\nu$ finite. Let us show that we may calculate it by the steepest descent method. Let $\phi=b\varphi$. Then
\eq$$
G_\nu({\{x_i\}_N})=b^{-N}Z_\nu^{-1}\int\cD\phi\,\phi(x_N)\cdots\phi(x_1)\e^{-b^{-2}(S_1[\phi]-\nu\phi(0))}.
\label{Gnu-S1}
$$
In the limit $b\to0$, the integral is determined by the stationary point:
$$
{\delta(S_1[\phi]-\nu\phi(0))\over\delta\phi(x)}=0,
$$
i.e.
\eq$$
\nabla^2\phi-m^2\sh\phi=-8\pi\nu\delta(x).
\label{shG-delta}
$$
Due to the rotational symmetry this equation admits a rotational symmetric solution, which only depends on $r=|x|$. It satisfies the radial sinh\-/Gordon equation
\eq$$
t^{-1}\d_t(t\,\d_t\phi)=\sh\phi,
\label{shG-radial}
$$%
as a function of $t=mr$. We will be interested in the solution that decays at infinity. At small values of $r$ we may neglect the potential term, and obtain the asymptotics
\eq$$
\phi(x)\simeq-4\nu\log r.
\label{phi-r-small}
$$
Here and below $r=|x|$, $r_i=|x_i|$ etc. In terms of this solution the classical limit of the correlation functions (\ref{Gnu-def}) reads
\eq$$
G_\nu(\{x_i\}_N)=b^{-N}\prod^N_{i=1}\phi(x_i)+O(b^{2-N}).
\label{Gnu-classical}
$$

In \cite{McCoy:1976cd} it was shown that there is a family of solutions $\phi_\nu(t)$ to equation (\ref{shG-radial}) that decay at infinity. These solutions are given explicitly by the series
\eq$$
\phi_\nu(t)
=\sum^\infty_{n=0}\lambda^{2n+1}\phi^{(2n+1)}(t),
\qquad
\lambda={\sin\pi\nu\over\pi},
\label{phinu-lambda-def}
$$
where (we give a simpler integral representation of~\cite{Cecotti:1992qh})
\eq$$
\phi^{(n)}(t)={4\over n}\int d^n\theta\,\prod^n_{i=1}{\e^{-t\ch\theta_i}\over2\ch{\theta_i-\theta_{i+1}\over2}}
\label{phi-r-theta}
$$
with $\theta_{n+1}=\theta_1$. The reason why the parameter $\nu$ is used instead of $\lambda$ will become clear below. Note that a real solution exists either for real values of $\nu$ or for complex values with $\Re\nu-{1\over2}\in\Z$. In this paper we will limit ourselves to real values of $\nu$.

Evidently,
\eq$$
\phi^{(n)}(t)={4\over n}\left(\pi\over2t\right)^{n/2}e^{-nt}\left(1+O\left(t^{-1}\right)\right),
\qquad
t\to\infty,
\label{phi-r-t-large}
$$
and, hence, $\phi^{(1)}(t)$ provides the leading asymptotics:
\eq$$
\phi_\nu(t)=\sqrt{8\pi}\lambda t^{-1/2}\e^{-t}\left(1+O\left(t^{-1}\right)\right),
\qquad
t\to\infty.
\label{phi-t-large}
$$

Note also~\cite{Zamolodchikov:1994uw} that the function
\eq$$
\psi_\nu(t)=\sum^\infty_{n=1}\lambda^{2n}\phi^{(2n)}(t)
\label{psinu-def}
$$
is the solution that decays at infinity to the companion equation
\eq$$
t^{-1}\d_t(t\,\d_t\psi)=\ch\phi-1
\label{psieq-radial}
$$
for $\phi=\phi_\nu$.

The small $t$ asymptotics of the function $\phi_\nu$ is given by~\cite{McCoy:1976cd}
\eq$$
\e^{\mp\phi_{\pm\nu}(t)/2}=\beta_\nu t^{2\nu}+\beta_{1-\nu}t^{2-2\nu}+O\left(t^2\right)
\quad(0\le\nu\le1),
\qquad
t\to0,
\label{phi-t-small}
$$
where
\eq$$
\beta_\nu=2^{-6\nu}{\Gamma({1\over2}-\nu)\over\Gamma({1\over2}+\nu)}.
\label{nu-B-def}
$$
Complete small $t$ expansions of the functions $\phi_\nu$, $\psi_\nu$ can be found in~\cite{Zamolodchikov:1994uw,Basor:1991ax,Gamayun:2013auu}. Comparing this asymptotics with (\ref{phi-r-small}) we see that
\eq$$
\phi(x)=\phi_\nu(mr)
\label{phi-phinu}
$$
for real values of $\nu$ in the region $|\nu|<{1\over2}$. For all other real values of $\nu$, the solution reduces to this region due to the reflection relations:
\eq$$
\phi_\nu(t)=\phi_{1-\nu}(t)=-\phi_{-\nu}(t),
\qquad
\psi_\nu(t)=\psi_{1-\nu}(t)=\psi_{-\nu}(t).
\label{phi-nu-reflection}
$$
We see that at the reflection points $\nu={1\over2}\bmod1$ the leading and the subleading contributions in the small distance asymptotics exchange their roles, which matches the periodicity in the parameter $\nu$ of the formulas (\ref{phinu-lambda-def}), (\ref{psinu-def}). The reflection relations (\ref{phi-nu-reflection}) are the classical counterparts of the quantum reflection relations~\cite{Zamolodchikov:1995aa,Fateev:1997nn}:
\eq$$
\e^{\alpha\varphi}=R_\alpha\e^{(Q-\alpha)\varphi}=R_{-\alpha}\e^{-(Q+\alpha)\varphi},
\qquad
Q=b^{-1}+b,
\label{exp-reflection}
$$
with a known reflection factor $R_\alpha$.

Now we are interested in the fluctuations around the stationary point:
\eq$$
\varphi(x)=b^{-1}\phi_\nu(mr)+\chi(x).
\label{chi-def}
$$
Define the regularized action
\Align$$
S^\reg_\nu
&=\left.
{1\over8\pi}\int_{r>\ve/m} d^2x\,\left(-{(\d_\mu\phi)^2\over2}+m^2(\ch\phi-1-\phi\sh\phi)\right)
-2\nu^2\log\ve\,\right|_{\substack{\phi(x)=\phi_\nu(mr)\\\ve\to0}}
\notag
\\
&=\left.{1\over4}\int^\infty_\ve dt\,t\left(-{\phi_\nu^{\prime\,2}(t)\over2}+\ch\phi_\nu(t)-1-\phi_\nu(t)\sh\phi_\nu(t)\right)
-2\nu^2\log\ve\,\right|_{\ve\to0}
\label{Sreg-nu-def}
$$
for $|\nu|<{1\over2}$. We have
\eq$$
S_b[\varphi]-b^{-1}\nu\varphi(0)=b^{-2}S^\reg_\nu+\Delta S[\chi]+\const\cdot\nu^2,
\label{shG-action-decomp}
$$
where the constant is infinite and
\Align$$
\Delta S[\chi]
&={1\over8\pi}\int d^2x\left({(\d_\mu\chi)^2\over2}+{m^2\over b^2}(\ch b\chi-1)\ch\phi_\nu(mr)
+{m^2\over b^2}(\sh b\chi-b\chi)\sh\phi_\nu(mr)\right)
\notag
\\
&={1\over8\pi}\int d^2x\left({(\d_\mu\chi)^2\over2}+{m^2\over2}\chi^2\ch\phi_\nu(mr)\right)
+S_\text{int}[\chi].
\label{shG-action-Delta}
$$
The interaction term can be expanded in powers of the field $\chi$:
\eq$$
S_\text{int}[\chi]=\sum^\infty_{n=3}S^{(n)}[\chi],
\qquad
S^{(n)}[\chi]={b^{n-2}m^2\over n!\,8\pi}\times\Cases{\int d^2x\,\chi^n\sh\phi_\nu(mr),&n\in2\Z+1;\\\int d^2x\,\chi^n\ch\phi_\nu(mr),&n\in2\Z.}
\label{S(n)-def}
$$
Already at this step we may conclude that due to eqs.~(\ref{phi-nu-reflection}) the reflection property holds in the semiclassical approximation:
\eq$$
G_\nu({\{x_i\}_N})=G_{1-\nu}({\{x_i\}_N})=(-1)^NG_{-\nu}({\{x_i\}_N}).
\label{Gnu-reflection}
$$
The renormalization of the reflection point (which means $\e^{b^{-1}\nu\varphi}\to\e^{(b^{-1}+b)\nu\varphi}$) and the overall reflection coefficient $R_\alpha$ are apparently nonperturbative effects.

The vacuum expectation value of the exponential operator $\cG_\nu$ was exactly found in~\cite{Lukyanov:1996jj}. In the semiclassical limit it is given by
\eq$$
\Gathered{
\cG_\nu
=\langle\e^{b^{-1}\nu\varphi(0)}\rangle
=D_\nu\left(m\over4\right)^{-2{\nu^2\over b^2}}
\exp\left(-{1\over b^2}\int^\infty_0{dt\over t}\left({\sh^22\nu t\over t\sh2t}-2\nu^2e^{-2t}\right)\right),
\\
D_\nu
=2^{-2\nu^2}\exp\left({1\over2}\int^\infty_0{dt\over t}{\sh^22\nu t\over\ch^2t}\right),
\qquad
|\nu|<{1\over2}.
}\label{Gnu-vac-def}
$$
Note that the vacuum expectation values are not analytic in $\nu$ at the points $\nu=\pm1/2$ in the semiclassical limit. In this paper we will consider all operators divided by $\cG_\nu$ like $V_\nu(x)$. The form factors of the  operators $V_\nu$ turn out to be analytical, while those for the descendant operators do not. We will discuss this non\-/analyticity in sec.~\ref{sec:form-factors-simplest}.

Since the Lagrangian is only $r$\-/dependent, it is natural to develop the perturbation theory on the basis of the radial quantization. Let
\eq$$
\chi(x)=\Q f_0(x)+4\i\P f_*(x)+\sum_{k\ne0}\left({\a_k\over\i k}f_k(x)+{\ba_k\over\i k}\bar f_k(x)\right),
\label{chi-reddecomp}
$$
where the operators satisfy the commutation relations
\eq$$
[\P,\Q]=-\i,
\qquad
[\a_k,\a_l]=[\ba_k,\ba_l]=2k\delta_{k+l},
\qquad
[\a_k,\ba_l]=[\a_k,\P]=\cdots=0.
\label{osc-commut}
$$
The radial ket vacuum $|0\rangle_\nu$ is defined as
\eq$$
\P|0\rangle_\nu=0,
\qquad
\a_k|0\rangle_\nu=\ba_k|0\rangle_\nu=0\quad(k>0).
\label{ketvac-def}
$$
It is convenient to introduce the state
\eq$$
|\alpha\rangle_\nu=\e^{\alpha\Q}|0\rangle_\nu=\e^{\alpha\chi(0)}|0\rangle_\nu,
\qquad
\P|\alpha\rangle_\nu=-\i\alpha|\alpha\rangle_\nu.
\label{nu-radstate-def}
$$
Define also a bra vacuum $\langle\infty|$ by the relations
\eq$$
\langle\infty|\Q=0,
\qquad
\langle\infty|\a_{-k}=\langle\infty|\ba_{-k}=0\quad(k>0),
\qquad
\langle\infty|0\rangle_\nu=1.
\label{bravac-def}
$$
Necessary consistency checks for these definitions are given in Appendix~\ref{app:consistency}.

Introduce the correlations functions of the quantum fluctuation:
\eq$$
\tilde G_\nu(\{x_i\}_N)=\langle\chi(x_N)\cdots\chi(x_1)V_\nu(0)\rangle
\sim\int\cD\chi\,\chi(x_N)\cdots\chi(x_1)\e^{-\Delta S[\chi]}.
\label{tildeG-def}
$$
Evidently,
\eq$$
G_\nu(\{x_i\}_N)=\sum_{I\subset\{1,\ldots,N\}}\prod_{i\in I}b^{-1}\phi_\nu(mr_i)\times\tilde G_\nu(\{x_i\>|\>i\not\in I\}).
\label{G-tildeG-rel}
$$
The correlation functions $\tilde G_\nu$ can be represented in the form
\eq$$
\tilde G_\nu(\{x_i\}_N)={\langle\infty|\T_r[\chi(x_N)\cdots\chi(x_1)\e^{-S_\text{int}}]|0\rangle_\nu\over\langle\infty|\T_r[\e^{-S_\text{int}}]|0\rangle_\nu},
\label{tildeG-matel}
$$
where $\T_r$ is the $r$ ordering, which is necessary in the radial quantization scheme since the variable $r=|x|$ plays here the role of time.

Note that from the fact
$$
\e^{\alpha\phi(x)}\e^{\beta\phi(0)}=|x|^{-4\alpha\beta}\e^{(\alpha+\beta)\phi(0)}
\quad\text{as $x\to0$},
$$
it follows that
$$
\e^{b\delta\chi(x)}V_\nu(0)\to{\cG_{\nu+b^2\delta}\over\cG_\nu(m^{2\nu}\beta_\nu)^{-2\delta}}V_{\nu+b^2\delta}
\quad\text{as $x\to0$ for $|\nu|<{1\over2}$.}
$$
Hence
\eq$$
{\langle\infty|\T_r[F[\varphi]\e^{-S_\text{int}}]|b\delta\rangle_\nu
  \over\langle\infty|\T_r[\e^{-S_\text{int}}]|b\delta\rangle_\nu}
={\cG_{\nu+b^2\delta}\over\cG_\nu(m^{2\nu}\beta_\nu)^{-2\delta}}
\times{\langle\infty|\T_r[F[\varphi]\e^{-S_\text{int}}]|0\rangle_{\nu+b^2\delta}
 \over\langle\infty|\T_r[\e^{-S_\text{int}}]|0\rangle_{\nu+b^2\delta}},
\label{deltaQ-nu-rel}
$$
where $F[\varphi]$ is any functional of the field $\varphi=b^{-1}\phi_\nu+\chi$. It is not difficult to check that
\eq$$
\lim_{b\to0}{\cG_{\nu+b^2\delta}\over\cG_\nu(m^{2\nu}\beta_\nu)^{-2\delta}}=1.
\label{Gration-one}
$$
Thus in the leading order in $b$ we may assume the equivalence
\eq$$
\e^{b\delta\Q}|0\rangle_\nu=|b\delta\rangle_\nu\leftrightarrow|0\rangle_{\nu+b^2\delta}
\quad\text{for $\delta\sim1$.}
\label{deltaQ-equivalence-b2}
$$
Moreover, up to a factor this equivalence is correct beyond the leading order. Later we will use it when studying the renormalization of operators.

The functions $f_*(x)$, $f_k(x)$, $\bar f_k(x)$ satisfy the equation
$$
\nabla^2f(x)=m^2f(x)\ch\phi_\nu(mr)
$$
and factorize as follows
\eq$$
f_*(x)=u_*(mr),
\qquad
f_k(x)=\e^{-\i k\xi}m^ku_k(mr),
\qquad
\bar f_k(x)=\e^{\i k\xi}m^ku_k(mr),
\label{f-fact}
$$
where $\xi$ is the polar angle:
$$
z=x^1+\i x^2=r\e^{\i\xi}.
$$
Here the functions $u_*(t)$, $u_k(t)$ have the following properties:
\eq$$
u_*(t)=-\log t+O(1),
\qquad
u_k(t)=t^{-k}\left(1+o\left(1\right)\right)
\quad
\text{as $t\to0$,}
\label{u-t-zero}
$$
and
\eq$$
u_*(t),u_k(t)=O\left(t^{-1/2}e^{-t}\right)
\quad
\text{as $t\to\infty$, if $k>0$.}
\label{u-t-infty}
$$
They satisfy the equation
\eq$$
u''+t^{-1}u'-(\ch\phi_\nu(t)+n^2t^{-2})u=0,
\label{Bessel-generalized}
$$
where $n=k$ for $u=u_k$ and $n=0$ for~$u=u_*$. The equation (\ref{Bessel-generalized}) reduces to the modified Bessel equation in the case $\nu=0$. We will refer to this generalization of the Bessel equation as the `sinh\-/Bessel equation'.

Recall the properties of the (modified) Bessel equation. For non\-/integer $n>0$ it has the following solutions
$$
I_{\pm n}(t)=\sum^\infty_{p=0}{1\over p!\,\Gamma(p\pm n+1)}\left(t\over2\right)^{2p\pm n}.
$$
Instead, we may use as a basis the solution $I_n(t)$ and the solution
$$
K_n(t)={\pi\over2\sin\pi n}(I_{-n}(t)-I_n(t)).
$$
The last basis remains non\-/degenerate for $n=k\in\Z$ as well. The functions $I_k$, $K_k$ have the following asymptotic properties:
\eq$$
\Aligned{
K_k(t)
&=2^{k-1}\,(k-1)!\,t^{-k}+O\left(t^{2-k}\log|t|\right),
\quad
&I_k(t)
&={t^k\over2^k\,k!}+O\left(t^{k+2}\right)
\quad(k>0),
\\
K_0(t)
&=-\log{t\over2}-\gamma_E+O\left(t^2\log t\right),
\quad
&I_0(t)
&=1+O\left(t^2\right)
}\label{KI-Bessel-t-small}
$$
as $t\to0$, where $\gamma_E$ is the Euler constant, and
\eq$$
K_k(t)=\sqrt{\pi\over2t}\e^{-t}\left(1+O\left(t^{-1}\right)\right),
\qquad
I_k(t)={1\over\sqrt{2\pi t}}\e^t\left(1+O\left(t^{-1}\right)\right)
\label{KI-Bessel-t-large}
$$
as $t\to\infty$, $|\Arg t|<{\pi\over2}$. After comparing with the asymptotic properties (\ref{u-t-zero}) and (\ref{u-t-infty}) we see that for $\nu=0$
\eq$$
u_*(t)=K_0(t),
\qquad
u_k(t)=\Cases{2^{1-k}\,(k-1)!^{-1}K_k(t),&k>0,\\
 2^{-k}\,(-k)!\,I_{-k}(t),&k\le0.}
\label{R-Bessel-final}
$$

Let us now generalize these functions to nonzero values of $\nu$. We will be only interested in integer values of $k$. Since $t^{-4|\nu|}\ll t^{-2}$ as $t\to0$ for $|\nu|<{1\over2}$, the leading small $t$ asymptotics of the functions $K_{\nu,k}(t)$, $I_{\nu,k}(t)$ remain unchanged for $\nu\ne0$. We normalize them in the same way as the Bessel functions:
\eq$$
K_{\nu,k}(t)=\Cases{2^{k-1}\,(k-1)!\,t^{-k}(1+o(1)),&k>0,\\-\log t+O(1),&k=0,}
\qquad
I_{\nu,k}(t)={t^k\over2^k\,k!}(1+o(1)),
\label{KIn-t-small-leading}
$$
so that
\eq$$
u_*(t)=K_{\nu,0}(t),
\qquad
u_k(t)=\Cases{2^{1-k}\,(k-1)!^{-1}K_{\nu,k}(t),&k>0,\\
 2^{-k}\,(-k)!\,I_{\nu,-k}(t),&k\le0.}
\label{u-KIn-final}
$$
More terms in the small $t$ expansions of these functions can be found in Appendix~\ref{app:smallt}.

Since $\ch\phi_\nu(t)-1=O\left(t^{-1}\e^{-2t}\right)$ as $t\to\infty$, the leading large distance asymptotics of $u(t)$ must be the same as that of the Bessel functions up to constant factors $K^\infty_{\nu,k}$, $I^\infty_{\nu,k}$:
\eq$$
K_{\nu,k}(t)=\sqrt{\pi\over2}t^{-1/2}\e^{-t}\left(K^\infty_{\nu,k}+O\left(t^{-1}\right)\right),
\qquad
I_{\nu,k}(t)={1\over\sqrt{2\pi}}t^{-1/2}\e^t\left(I^\infty_{\nu,k}+O\left(t^{-1}\right)\right).
\label{KI-t-large}
$$
We will also assume by definition
\eq$$
K_{\nu,-k}(t)=K_{\nu,k}(t),
\qquad
I_{\nu,-k}(t)=I_{\nu,k}(t).
\label{KInu-keven}
$$

The coefficients $K^\infty_{\nu,k}$, $I^\infty_{\nu,k}$ of relative normalization of the asymptotics of solutions of (\ref{Bessel-generalized}) as $t\to0$ and $t\to\infty$ (connection coefficients) are crucial for calculation of the form factors of descendant operators in the semiclassical limit. In fact, these two types of coefficients are related. Indeed, the functions $I_{\nu,k}(t)$ can be expressed in terms of the functions $K_{\nu,k}$ by means of the Wronskian. Let $f(t)$ and $g(t)$ be arbitrary solutions to the generalized Bessel equation~(\ref{Bessel-generalized}). It is easy to show that
\eq$$
fg'-gf'=Ct^{-1}
\label{Besgen-Wronskian}
$$
with a constant $C$. One may solve this equation with respect to $g(t)$:
\eq$$
g(t)=Cf(t)\int{dt\over tf^2(t)}.
\label{g-f-equation}
$$
Now choose $f(t)=K_{\nu,k}(t)$, $g(t)=I_{\nu,k}(t)$. Substituting the small distance asymptotics (\ref{KIn-t-small-leading}) into (\ref{Besgen-Wronskian}), we obtain $C=1$ for all $k\ge0$. Hence,
\eq$$
I_{\nu,k}(t)=K_{\nu,k}(t)\int^t_0{ds\over sK^2_{\nu,k}(s)}.
\label{IK-relation}
$$
Now consider the large $t$ asymptotics (\ref{KI-t-large}). Substituting them into the left and right hand sides of (\ref{IK-relation}) we immediately obtain
\eq$$
I^\infty_{\nu,k}K^\infty_{\nu,k}=1.
\label{IK-infty-relation}
$$

By taking the $\nu$ and $t$ derivatives of the sinh-Gordon equation, we easily obtain
\AlignAt{2}$$
K_{\nu,0}(t)
&={1\over4}{\d\phi_\nu(t)\over\d\nu},
\qquad
&K^\infty_{\nu,0}
&=\cos\pi\nu
\label{K0-phi-rel}
\\
K_{\nu,1}(t)
&=-{1\over4\nu}\phi'_\nu(t),
&K^\infty_{\nu,1}
&={\sin\pi\nu\over\pi\nu}
\qquad
(|\nu|<\thalf).
\label{K1-phi-rel}
$$
The calculation of other normalization constants is much less straightforward. In the next section we will prove the following two statements:

\begin{statement}\label{statement-Kinf}For $k\ge0$, the connection coefficients $K_{\nu,k}^\infty$ are given by
\eq$$
K^\infty_{\nu,k}
={\Gamma^2\left(k+1\over2\right)\over\Gamma\left({k+1\over2}+\nu\right)\Gamma\left({k+1\over2}-\nu\right)}
=\prod_{0\le i<{k-1\over2}}\left(1-{\nu^2\over\left({k-1\over2}-i\right)^2}\right)^{-1}
\times\Cases{\cos\pi\nu,&k\in2\Z,\\(\pi\nu)^{-1}\sin\pi\nu,&k\in2\Z+1.}
\label{Knu-infty-final}
$$
We will also adopt the convention that $K^\infty_{\nu,-k}=K^\infty_{\nu,k}$.
\end{statement}

\begin{statement}\label{statement-Kdecomp}The functions $K_{\nu,k}(t)$ are given by
\eq$$
{K_{\nu,k}(t)\over K^\infty_{\nu,k}}
\equiv U_k(t)=\sum^\infty_{n=0}\lambda^{2n}U^{(2n+1)}_k(t),
\qquad
U^{(n)}_k(t)=\int d^n\theta\,\ch k\theta_1\prod^n_{i=1}{\e^{-t\ch\theta_i}\over2\ch{\theta_i-\theta_{i+1}\over2}}.
\label{Knu-final}
$$
\end{statement}

In what follows, for calculations involving the cubic interaction term $S^{(3)}$ we will need the asymptotics of the integrals
\eq$$
\cI^\pm_{kl}(t_0)=\int^\infty_{t_0}dt\,tI_{\nu,|k\pm l|}(t)K_{\nu,k}(t)K_{\nu,l}(t)\sh\phi_\nu(t)
\label{cI-pm-def}
$$
and
\eq$$
\cI^0_{kl}(t_0)=\int^{t_0}_0dt\,tI_{\nu,k-l}(t)K_{\nu,k}(t)I_{\nu,l}(t)\sh\phi_\nu(t)
\label{cI-0-def}
$$
for small values of $t_0$. Let us summarize some of their properties to be used later.

The integrals $\cI^0_{kl}$ are straightforwardly calculated from the small $t$ asymptotics of the generalized Bessel functions. The leading terms are
\eq$$
\cI^0_{kl}(t_0)={(k-1)!\over8\,(k-l)!\,l!}\left({\beta_\nu^{-2}t_0^{2-4\nu}\over1-2\nu}-{\beta_\nu^2t_0^{2+4\nu}\over1+2\nu}\right)+O\left(t_0^{4-8|\nu|}\right).
\label{cI-0-explicit}
$$
for $|\nu|<1/2$. We will also need subleading terms in this expansion for $k=l=2$:
\Align$$
\cI^0_{22}(t_0)
&={\beta_\nu^{-2}t_0^{2-4\nu}\over16(1-2\nu)}-{\beta_\nu^2t_0^{2+4\nu}\over16(1+2\nu)}
+{\beta_\nu^{-4}(1+4\nu-4\nu^2)t_0^{4-8\nu}\over64(1-2\nu)^3(1+2\nu)(3-2\nu)}
-{\beta_\nu^4(1-4\nu-4\nu^2)t_0^{4+8\nu}\over64(1+2\nu)^3(1-2\nu)(3+2\nu)}
\notag
\\
&\quad
-{\nu t^4\over16(1-4\nu^2)(9-4\nu^2)}+O(t_0^{6-12|\nu|}).
\label{cI-0-k=l=2}
$$

The integrals $\cI^\pm_{kl}$ are taken over a large range of the variable $t$ and cannot be reduced to the asymptotics directly. Nevertheless, we will see in the next section that the corresponding indefinite integrals can be evaluated in the form of special combinations of the generalized Bessel functions. Thus the integrals $\cI^\pm_{kl}$ are expressed in terms of the asymptotics of these combinations. The functions $\cI^+_{kl}(t_0)$ have a finite limit as $t_0\to0$, so that we can define
\eq$$
\cV^+_{kl}=\cV^+_{lk}=\cI^+_{kl}(0).
\label{cV-p-def}
$$
These limiting values are given by
\eq$$
\cV^+_{kl}={K^\infty_{\nu,k}K^\infty_{\nu,l}\over4\lambda}\left({1\over K^\infty_{\nu,kl}}-{1\over K^\infty_{\nu,k+l}}\right).
\label{cVp-kl-fin}
$$
where
\eq$$
K^\infty_{\nu,kl}={2(-1)^{k+l}\,(k+l-1)!\,\lambda\over A_{kl}}.
\label{Kinf-kl-def}
$$
The numbers $A_{kl}$ are defined by the recursion relations (\ref{BA-rec}) with the initial condition (\ref{A0l-A1l-explicit}). For $l=1$ due to $K^\infty_{\nu,k1}=K^\infty_{\nu,k}$ we have explicitly
\eq$$
\cV^+_{k1}={1\over4\nu}\left(1-{K^\infty_{\nu,k}\over K^\infty_{\nu,k+1}}\right).
\label{cVp-k1-fin}
$$

The integrals $\cI^-_{kl}(t_0)$ diverge as $t_0\to0$. The leading term can be easily found from the asymptotics (\ref{KIn-t-small-leading}), (\ref{phi-t-small}):
\eq$$
\cI^-_{kl}(t_0)=2^{2l-4}{(k-1)!\,(l-1)!\over(k-l)!}\left({\beta_\nu^{-2}t_0^{-2l+2-4\nu}\over l-1+2\nu}-{\beta_\nu^2t_0^{-2l+2+4\nu}\over l-1-2\nu}\right)
+O(t_0^{-2l+4-8|\nu|}\log t_0)
\quad(k\ge l).
\label{cIm-firstterm}
$$
In fact, we need expansions of these integrals up to the finite terms. We have no general expression for these expansions, but there are explicit expressions available in two special cases. First, for $l=1$ we have
\eq$$
\cI^-_{k1}={\beta_\nu^{-2}t_0^{-4\nu}+\beta_\nu^2t_0^{4\nu}\over8\nu}+\cV^-_{k1}+O(t_0^{2-8|\nu|}),
\qquad
\cV^-_{k1}=-{1\over4\nu}{K^\infty_{\nu,k}\over K^\infty_{\nu,k-1}}.
\label{cIm-k1-fin}
$$
Second, for $k=l=2$ we have
\eq$$
\Aligned{
\cI^-_{22}(t_0)
&={\beta_\nu^{-2}t_0^{-2-4\nu}\over1+2\nu}-{\beta_\nu^2t_0^{-2+4\nu}\over1-2\nu}
+{\beta_\nu^{-4}t_0^{-8\nu}\over4(1-2\nu)^2(1+2\nu)}-{\beta_\nu^4t_0^{8\nu}\over4(1+2\nu)^2(1-2\nu)}+\cV^-_{22}+O(t_0^{2-12|\nu|}),
\\
\cV^-_{22}
&={\nu\over(1-4\nu^2)^2}\left(1+{\pi\tg\pi\nu\over4\nu}\right).
}\label{cIm-22-fin}
$$
We see that these formulas provide correct decomposition in some narrower region of $\nu$: $|\nu|<{1\over4}$ for $\cI^-_{k1}$ and $|\nu|<{1\over6}$ for $\cI^-_{22}$. Nevertheless, it will allow us to establish renormalized operators in some finite vicinity of the point $\nu=0$. It seems plausible that the resulting correlation functions and form factors for the renormalized operators can be analytically continued to the entire region $|\nu|<{1\over2}$.

Note that the functions introduced in this section have simpler and more transparent analogs in the case of the Liouville theory. We discuss them in Appendix~\ref{app:Liouville}.

\section{Generalized Bessel functions and Fredholm determinants}
\label{sec:fredholm-dets}

\subsection{Tracy\--Widom approach}

It turns out that the most natural and effective way of studying the functions introduced in the last section is to employ the Fredholm determinant approach of Tracy and Widom~\cite{Tracy:1996CMP179}. Let us recall some of the relevant results. Consider the space of functions $L^2(0,\infty)$ with the scalar product
\eq$$
\langle f|g\rangle=\int^\infty_0dx\,f(x)g(x).
\label{fs-scalar}
$$
Consider an integral operator $\hK$ on $L^2(0,\infty)$ with the kernel
\eq$$
K(x,y)={E(x)E(y)\over x+y}.
\label{kernelK}
$$
The function $E(x)$ can be very general. Essentially, all we need is to ensure that the Fredholm determinants $\det(1\pm\hK)$ are well defined. We will choose this function in the form
\eq$$
E(x)=\exp{1\over2}\sum_{k\in\Z}t_kx^k.
\label{E(x)-gen}
$$
The statement is that the functions defined by
\eq$$
\Aligned{
\thalf\phi(\{t_k\})
&=\log\det(1+\hK)-\log\det(1-\hK),
\\
\thalf\psi(\{t_k\})
&=-\log\det(1+\hK)-\log\det(1-\hK),
}\label{phi-Fd-def}
$$
satisfy the sinh\-/Gordon equation and its `companion' equation:
\eq$$
\d_1\d_{-1}\phi=\sh\phi,
\qquad
\d_1\d_{-1}\psi=\ch\phi-1,
\label{shGeq-gen}
$$
where $\d_k=\d/\d t_k$. It immediately identifies
\eq$$
t_1=-mz/2,
\qquad
t_{-1}=-m\bz/2.
\label{tpm1-z-id}
$$
The choice of the signs is somewhat nonstandard to simplify some formulas below. The function $\phi$ as a function of $t_1$ as a space variable satisfies the mKdV integrable hierarchy in the times $t_3,t_5,t_7,\ldots$, and the sinh\-/Gordon hierarchy in the times $t_{-1},t_{-3},t_{-5}\ldots$ . The even variables $t_{2k}$ do not have such a transparent interpretation.

The equations (\ref{shG-radial}), (\ref{psieq-radial}) are obtained by assuming
\eq$$
t_k=0\ \text{for $|k|\ge2$,}
\qquad
t_{\pm1}=-{t\over2}\e^{\pm\i\xi},
\qquad
t_0=\log\lambda.
\label{basic-reduction}
$$
We will call this assumption the basic reduction. For any function $f(\{t_k\})$ its basic reduction will be referred to as $f^\text{r}$ and will be considered as a function either of $t_{\pm1}$ or of $t,\xi$, which depends on $\lambda$ (or $\nu$) as on a parameter. In the basic reduction the functions $\phi$ and $\psi$ are $\xi$\-/independent for $|\Re\xi|<\pi$ and $t>0$ and coincide with the functions defined in (\ref{phinu-lambda-def}), (\ref{psinu-def}):
\eq$$
\phi^\text{r}=\phi_\nu(t),
\qquad
\psi^\text{r}=\psi_\nu(t).
\label{phi-psi-identification}
$$

The expansions (\ref{phinu-lambda-def}), (\ref{psinu-def}) with (\ref{phi-r-theta}) are obtained straightforwardly from the definition (\ref{phi-Fd-def}) by expanding it in $\hK$ with the substitution $x_i=\e^{\theta_i}$ for the integration variables. It is important to note that though the large $t$ expansion of $\phi$ and $\psi$ is known for general values of $t_k$, the small $t$ expansion is only known in the basic reduction.

Now let us return to the general values of the variables $t_k$. Denote
\eq$$
\hU={1\over1-\hK^2},
\qquad
\hV={\hK\over1-\hK^2}.
\label{cUcV-def}
$$
In~\cite{Tracy:1996CMP179} the equations (\ref{shGeq-gen}) were proved by introducing scalar products
\eq$$
u_{i,j}=u_{j,i}=\langle E_i|\hU|E_j\rangle,
\qquad
v_{i,j}=v_{j,i}=\langle E_i|\hV|E_j\rangle,
\qquad
i,j\in\Z,
\label{uij-vij-def}
$$
where $E_i(x)=x^iE(x)$, and showing that they satisfy certain differentiation and recursion formulas. We do not need all of these relations here, so let us only record a few important consequences:
\Gather$$
\d_1\phi=2u_{0,0},
\qquad
\d_{-1}\phi=2u_{-1,-1},
\qquad
\d_1\psi=2v_{0,0},
\qquad
\d_{-1}\psi=2v_{-1,-1},
\label{phi-psi-u-v-rel}
\\
2\,\d_1^2\psi=(\d_1\phi)^2,
\qquad
2\,\d_{-1}^2\psi=(\d_{-1}\phi)^2.
\label{psi-second-derivs}
$$

\subsection{Generalized Bessel functions and recursion relations}

Let us define the functions
\eq$$
\Phi_{k_1k_2\ldots k_s}=\d_{k_1}\d_{k_2}\cdots\d_{k_s}\phi,
\qquad
\Psi_{k_1k_2\ldots k_s}=\d_{k_1}\d_{k_2}\cdots\d_{k_s}\psi.
\label{Phi-Psi-func-def}
$$
Below we will study in detail the properties of these functions for $s=1,2$. Start with the case $s=1$. By taking the $t_k$ derivative of the equations (\ref{shGeq-gen}) we obtain
\eq$$
(\d_1\d_{-1}-\ch\phi)\Phi_k=0,
\qquad
\d_1\d_{-1}\Psi_k=\Phi_k\sh\phi.
\label{Phik-Psik-eqs}
$$

Let us rewrite the function $\Phi_k$ in terms of the kernel (\ref{kernelK}). Define the operator $\hM$ of the multiplication by~$x$, so that
\eq$$
\hM|E_i\rangle=|E_{i+1}\rangle.
\label{Mop-def}
$$
From this we easily derive
\eq$$
\d_k\hK={1\over2}(\hM^k\hK+\hK \hM^k).
\label{dk-K}
$$
Then by taking the $t_k$ derivative of the definition (\ref{phi-Fd-def}) by means of (\ref{dk-K}) and using the cyclic property of the trace we obtain
\eq$$
\Phi_k=4\tr\left(\hM^k\hV\right),
\qquad
\Psi_k=4\tr\left(\hM^k(\hU-1)\right).
\label{Phik-Psik-tr}
$$

Now in the basic reduction (\ref{basic-reduction}) we have
\eq$$
\Phi_k^\text{r}=\tilde\Phi_k(t)\e^{-\i k\xi},
\qquad
\Psi_k^\text{r}=\tilde\Psi_k(t)\e^{-\i k\xi}.
\label{Phi-Psi-rad}
$$
By substituting this into (\ref{Phik-Psik-eqs}) we obtain the equations
\eq$$
\Gathered{
\tilde\Phi_k''+t^{-1}\tilde\Phi_k'-(\ch\phi_\nu+k^2t^{-2})\tilde\Phi_k=0,
\\
\tilde\Psi_k''+t^{-1}\tilde\Psi_k'-k^2t^{-2}\tilde\Psi_k=\tilde\Phi_k\sh\phi_\nu.
}\label{tilde-Phi-Psi-eqs}
$$
The first equation is nothing but the generalized Bessel equation (\ref{Bessel-generalized}). As $t\to\infty$ we have
\eq$$
\tilde\Phi_k(t)\simeq4\tr(\hM^k\hK)\Bigr|_{\substack{t_1=t_{-1}\\=-t/2}}=2\lambda\int^\infty_0dx\,x^{k-1}\e^{-{t\over2}(x+x^{-1})t}
=2\lambda\int^\infty_{-\infty}d\theta\,\e^{k\theta-t\ch\theta}\simeq\sqrt{8\pi}\lambda t^{-1/2}\e^{-t}.
\label{Phik-t-large}
$$
Up to a factor the only solution to equation (\ref{Bessel-generalized}) that decreases at infinity is $K_{\nu,k}$. Comparing this asymptotics with (\ref{KI-t-large}) we obtain
\eq$$
\tilde\Phi_k(t)=4\lambda U_k(t)=4\lambda{K_{\nu,k}(t)\over K^\infty_{\nu,k}}.
\label{Phik-Knuk-rel}
$$
In particular, we see that equations (\ref{K0-phi-rel}), (\ref{K1-phi-rel}) are special cases of (\ref{Phi-Psi-func-def}). Statement~\ref{statement-Kdecomp} immediately follows from the expansion of the first equation of (\ref{Phik-Psik-tr}) in the series in $\hK^2$.

To prove Statement~\ref{statement-Kinf}, we will need the small $t$ expansion of $\tilde\Phi_k$, however the small $t$ expansion of $\phi$ is only known for $t_k=0$ ($|k|\ge2$). To circumvent this difficulty, let us derive recursion relations for $\Phi_k$ and~$\Psi_k$.

\begin{theorem}\label{theorem-Phik-Psik-recursion}
We have the following differential formulas
\subeq{\label{Phik-Psik-recursion}
\Align$$
\d_1^2\Psi_k
&=\Phi_1\d_1\Phi_k,
\tag{\ref{Phik-Psik-recursion}a}
\\
\d_{-1}^2\Psi_k
&=\Phi_{-1}\d_{-1}\Phi_k,
\tag{\ref{Phik-Psik-recursion}b}
\\
\d_1^2\Phi_k
&=\Phi_{k+2}+\Phi_1\d_1\Psi_k,
\tag{\ref{Phik-Psik-recursion}c}
\\
\d_{-1}^2\Phi_k
&=\Phi_{k-2}+\Phi_{-1}\d_{-1}\Psi_k.
\tag{\ref{Phik-Psik-recursion}d}
$$}
\end{theorem}
Equations (\ref{Phik-Psik-recursion}a,b) immediately follow from equations (\ref{psi-second-derivs}) by applying $\d_k$. The proof of (\ref{Phik-Psik-recursion}c,d) is more complicated and goes as follows.

First of all, note that the definition (\ref{kernelK}) can be rewritten as
\eq$$
\hM\hK+\hK \hM=|E_0\rangle\langle E_0|.
\label{MK-anticommut}
$$
Hence,
\subeq{\label{MUV-commut}
\Align$$
\hM\hV+\hV\hM
&=\hU|E_0\rangle\langle E_0|\hU-\hV|E_0\rangle\langle E_0|\hV,
\tag{\ref{MUV-commut}a}
\\
\hM\hU-\hU\hM
&=\hU|E_0\rangle\langle E_0|\hV-\hV|E_0\rangle\langle E_0|\hU.
\tag{\ref{MUV-commut}b}
$$}

Since $\Phi_k=2\tr \hM^{k-1}(\hM\hV+\hV \hM)$ (cf the first equation in (\ref{Phik-Psik-tr})), the identity (\ref{MUV-commut}a) implies that
\eq$$
\Phi_k=2\langle E_0|\hU \hM^{k-1}\hU|E_0\rangle-2\langle E_0|\hV \hM^{k-1}\hV|E_0\rangle.
\label{Phik-alternative}
$$

From (\ref{dk-K}) and (\ref{MK-anticommut}) we have
\eq$$
\d_1\hK={1\over2}|E_0\rangle\langle E_0|,
\qquad
\d_{-1}\hK={1\over2}|E_{-1}\rangle\langle E_{-1}|.
\label{dpm1-K}
$$
Using the general formulas for differentiating
$$
\Aligned{
d\hU
&=\hV\,d\hK\,\hU+\hU\,d\hK\,\hV,
\\
d\hV
&=\hU\,d\hK\,\hU+\hV\,d\hK\,\hV,
}
$$
we obtain
\subeq{\label{dpm1-UV}
\Align$$
2\,\d_1\hU
&=\hV|E_0\rangle\langle E_0|\hU+\hU|E_0\rangle\langle E_0|\hV,
\tag{\ref{dpm1-UV}a}
\\
2\,\d_1\hV
&=\hU|E_0\rangle\langle E_0|\hU+\hV|E_0\rangle\langle E_0|\hV,
\tag{\ref{dpm1-UV}b}
\\
2\,\d_{-1}\hU
&=\hV|E_{-1}\rangle\langle E_{-1}|\hU+\hU|E_{-1}\rangle\langle E_{-1}|\hV,
\tag{\ref{dpm1-UV}c}
\\
2\,\d_{-1}\hV
&=\hU|E_{-1}\rangle\langle E_{-1}|\hU+\hV|E_{-1}\rangle\langle E_{-1}|\hV.
\tag{\ref{dpm1-UV}d}
$$
}
From (\ref{dpm1-UV}a,b) and (\ref{Phik-Psik-tr}) it follows that
\subeq{\label{d1-Phik-Psik}
\Align$$
\d_1\Phi_k
&=2\langle E_0|\hU\hM^k\hU|E_0\rangle+2\langle E_0|\hV\hM^k\hV|E_0\rangle,
\tag{\ref{d1-Phik-Psik}a}
\\
\d_1\Psi_k
&=2\langle E_0|\hU\hM^k\hV|E_0\rangle+2\langle E_0|\hV\hM^k\hU|E_0\rangle.
\tag{\ref{d1-Phik-Psik}b}
$$}
To take the second $\d_1$ derivative, we use the obvious identities
\eq$$
\d_k|E_i\rangle={1\over2}|E_{i+k}\rangle,
\qquad
\d_k\langle E_i|={1\over2}\langle E_{i+k}|.
\label{dk-Ei}
$$
Using (\ref{dk-Ei}) and (\ref{dpm1-UV}a,b), we obtain
\Align*$$
\d_1^2\Phi_k
&=\langle E_1|\hU\hM^k\hU|E_0\rangle+\langle E_0|\hU\hM^k\hU|E_1\rangle+\langle E_1|\hV\hM^k\hV|E_0\rangle+\langle E_0|\hV\hM^k\hV|E_1\rangle
\\*
&\quad+2u_{0,0}\left(\langle E_0|\hV\hM^k\hU|E_0\rangle+\langle E_0|\hU\hM^k\hV|E_0\rangle\right)
+2v_{0,0}\left(\langle E_0|\hU\hM^k\hU|E_0\rangle+\langle E_0|\hV\hM^k\hV|E_0\rangle\right).
$$
In the first line, we apply the commutation relations (\ref{MUV-commut}). Moving $\hM$ to the centers of expressions provides two terms that contain $\hM^{k+1}$ and combine into $\Phi_{k+2}$ according to (\ref{Phik-alternative}). The terms coming from the r.h.s.\ of (\ref{MUV-commut}) cancel the contribution proportional to $v_{0,0}$ and double the contribution proportional to $u_{0,0}=\Phi_1/2$. This gives the term of the form $\Phi_1\,\d_1\Psi_k$. This proves (\ref{Phik-Psik-recursion}c). The relation (\ref{Phik-Psik-recursion}d) immediately follows from the symmetry $t_k\to t_{-k}$.

Now let us study the asymptotics of the functions $\Phi_k$ as $t_{\pm1}\to0$ and $t_{-1}\sim t_1$ for $t_l=0$ ($|l|\ge2$). Assume also that $|\nu|<1/2$. For the sake of definiteness let $k\ge0$. Then from (\ref{phi-t-small}) we obtain $\Phi_0^\text{r}=-A_0\log t_1t_{-1}+o(\log t_1)$ and $\Phi_1^\text{r}=-A_1t_1^{-1}+o(t_1^{-1})$ with
\eq$$
A_0={2\tg\pi\nu\over\pi},
\qquad
A_1=-2\nu.
\label{A0-A1}
$$
From (\ref{psi-second-derivs}) we obtain $\Psi_0=-B_0\log t_1t_{-1}+o(\log t_1)$ and $\Psi_1=-B_1t_1^{-1}+o(t_1^{-1})$ with
\eq$$
B_0={4\nu\tg\pi\nu\over\pi},
\qquad
B_1=-2\nu^2 t_1^{-1}.
\label{B0-B1}
$$
By counting powers in (\ref{Phik-Psik-recursion}a,c) we get generally
\eq$$
\Aligned{
\Phi_0^\text{r}
&=-A_0\log t_1t_{-1}+o(\log t_1),
&\Phi_k^\text{r}
&=A_kt_1^{-k}+o(t_1^{-k}),
\\
\Psi_0^\text{r}
&=-B_0\log t_1t_{-1}+o(\log t_1),
&\Psi_k^\text{r}
&=B_kt_1^{-k}+o(t_1^{-k}).
}\label{Phik-Psik-t-small}
$$
Moreover, the identities (\ref{Phik-Psik-recursion}a,c) provide relations between the coefficients:
\subeq{\label{Ak-Bk-recursion}
\Align$$
B_k
&={2\nu\over k+1}A_k,
\tag{\ref{Ak-Bk-recursion}a}
\\
A_{k+2}
&=\bar k((k+1)A_k-A_1B_k)
=\bar k(k+1)\left(1-{4\nu^2\over(k+1)^2}\right)A_k.
\tag{\ref{Ak-Bk-recursion}b}
$$}
Here $\bar k=\max(1,k)$. These relations are easily solved with
\eq$$
A_k=(\overline{k-1})!\prod_{0\le i<{k-1\over2}}\left(1-{\nu^2\over\left({k-1\over2}-i\right)^2}\right)
\times\Cases{A_0,&k\in2\Z,\\A_1,&k\in2\Z+1.}
\label{Ak-final}
$$
Now from the relation (\ref{Phik-Knuk-rel}) and the asymptotics (\ref{KIn-t-small-leading}) we obtain
\eq$$
K^\infty_{\nu,k}=(-1)^k\,(\overline{k-1})!\,{2\lambda\over A_k},
\qquad
k\ge0.
\label{Kinfty-Ak-rel}
$$
This proves Statement~\ref{statement-Kinf}.

The differential formulas (\ref{Phik-Psik-recursion}) can be used to obtain first terms of small $t$ expansions for the functions $\Phi_k(t)$, $\Psi_k(t)$ and, hence, $K_{\nu,k}(t)$ by recursion in $k$. The base of the recursion is the expansions for the functions $\Phi_k(t)$, $\Psi_k(t)$ with $k=0,1$, which can be obtained directly from the known expansions for $\phi_\nu(t)$ and $\psi_\nu(t)$. For this purpose it is convenient to use the expansions in the form of~\cite{Zamolodchikov:1994uw}.

\subsection{Integrals of generalized Bessel functions}

Now let us study the functions $\Phi_{kl}$, $\Psi_{kl}$. Differentiating the equations (\ref{Phik-Psik-eqs}) with respect to $t_l$, we obtain
\subeq{\label{Phikl-Psikl-eqs}
\Align$$
(\d_1\d_{-1}-\ch\phi)\Phi_{kl}
&=\Phi_k\Phi_l\sh\phi,
\tag{\ref{Phikl-Psikl-eqs}a}
\\
\d_1\d_{-1}\Psi_{kl}-\Phi_{kl}\sh\phi
&=\Phi_k\Phi_l\ch\phi.
\tag{\ref{Phikl-Psikl-eqs}b}
$$}
In fact, we will only need the first equation, while the second one is given for completeness. In terms of the kernel $\hat K$, these functions are given by
\eq$$
\Phi_{kl}=4\sum^\infty_{n=1}\sum^{2n-1}_{m=1}\tr(\hM^k\hK^{m-1}\hM^l\hK^{2n-m}),
\qquad
\Psi_{kl}=-4\sum^\infty_{n=1}\sum^{2n}_{m=1}\tr(\hM^k\hK^{m-1}\hM^l\hK^{2n+1-m}).
\label{Phikl-Psikl-tr}
$$

In the basic reduction (\ref{basic-reduction}) we have
\eq$$
\Phi_{kl}^\text{r}=\tilde\Phi_{kl}(t)\e^{-\i(k+l)\xi},
\qquad
\Psi_{kl}^\text{r}=\tilde\Psi_{kl}(t)\e^{-\i(k+l)\xi}.
\label{Phikl-Psikl-rad}
$$
The equations for the radial parts reduce to
\subeq{\label{Phikl-Psikl-rad-eqs}
\Align$$
(\d_t^2+t^{-1}\d_t-\ch\phi_\nu-(k+l)^2t^{-2})\tilde\Phi_{kl}(t)
&=\tilde\Phi_k(t)\tilde\Phi_l(t)\sh\phi_\nu,
\tag{\ref{Phikl-Psikl-rad-eqs}a}
\\
(\d_t^2+t^{-1}\d_t-(k+l)^2t^{-2})\tilde\Psi_{kl}(t)-\tilde\Phi_{kl}(t)\sh\phi_\nu
&=\tilde\Phi_k(t)\tilde\Phi_l(t)\ch\phi_\nu.
\tag{\ref{Phikl-Psikl-rad-eqs}b}
$$}
Define the function $K_{\nu,kl}(t)$ by
\eq$$
\tilde\Phi_{kl}(t)={16\lambda^2K_{\nu,kl}(t)\over K^\infty_{\nu,k}K^\infty_{\nu,l}}.
\label{Knukl-def}
$$
Then eq.~(\ref{Phikl-Psikl-rad-eqs}a) reads
\eq$$
(\d_t^2+t^{-1}\d_t-\ch\phi_\nu-(k+l)^2t^{-2})K_{\nu,kl}(t)=K_{\nu,k}(t)K_{\nu,l}(t)\sh\phi_\nu.
\label{Knukl-eq}
$$
Recall the convention (\ref{KInu-keven}), so that $k$ and $l$ are not supposed to be necessarily positive. Multiplying the previous equation by $tI_{\nu,k+l}(t)$ and subtracting the equation for $I_{\nu,k+l}(t)$ multiplied by $tK_{\nu,kl}(t)$, we obtain a total derivative in the l.h.s. Then integrating by $t$ yields
\eq$$
W_{kl}(t)\equiv t(I_{\nu,k+l}(t)K_{\nu,kl}'(t)-I_{\nu,k+l}'(t)K_{\nu,kl}(t))=\int dt\,tI_{\nu,k+l}(t)K_{\nu,k}(t)K_{\nu,l}(t)\sh\phi_\nu(t).
\label{IKkl-Wronskian}
$$
Thus the integrals
\eq$$
\cI^+_{kl}(t_0)=W_{kl}(t)\Bigr|^\infty_{t=t_0},
\qquad
\cI^-_{kl}(t_0)=W_{k,-l}(t)\Bigr|^\infty_{t=t_0},
\qquad
k\ge l\ge0,
\label{cVpm-Wronskian}
$$
are governed by the asymptotics of the functions $I_{\nu,k\pm l}$ and $K_{\nu,k,\pm l}$. Due to (\ref{Phikl-Psikl-tr}) the large $t$ asymptotics of $\tilde\Phi_{kl}(t)$ is given by the same formula (\ref{Phik-t-large}) as $\tilde\Phi_k$. Thus the large $t$ asymptotics of $W_{kl}$ is simple,
\eq$$
W_{kl}(t)\to-{K^\infty_{\nu,k}K^\infty_{\nu,l}\over4\lambda K^\infty_{\nu,k+l}}
\quad\text{as $t\to\infty$.}
\label{Wkl-t-large}
$$

It is not as easy to find the small $t$ asymptotics. To do it we will need a system of recursion relations. By taking the $t_k$ derivative of the relations (\ref{Phik-Psik-recursion}) we get
\subeq{\label{Phikl-Psikl-recursion}
\Align$$
\d_1^2\Psi_{kl}
&=\d_1\Phi_k\,\d_1\Phi_l+\Phi_1\,\d_1\Phi_{kl},
\\
\d_{-1}^2\Psi_{kl}
&=\d_{-1}\Phi_k\,\d_{-1}\Phi_l+\Phi_{-1}\,\d_{-1}\Phi_{kl},
\\
\d_1^2\Phi_{kl}
&=\Phi_{k+2,l}+\d_1\Phi_l\,\d_1\Psi_k+\Phi_1\,\d_1\Psi_{kl},
\\
\d_{-1}^2\Phi_{kl}
&=\Phi_{k-2,l}+\d_{-1}\Phi_l\,\d_{-1}\Psi_k+\Phi_{-1}\,\d_{-1}\Psi_{kl}.
$$}

Let us find first the small $t$ asymptotics of $\Phi_{kl}^\text{r}$, $\Psi_{kl}^\text{r}$ for $k,l\ge0$. It is straightforward to show that the leading asymptotics are given by
\eq$$
\Aligned{
\Phi_{00}^\text{r}
&=-A_{00}\log t_1t_{-1}+o(\log t_1),
&\quad
\Phi_{kl}^\text{r}
&=A_{kl}t_1^{-k-l}+o(t_1^{-k-l})\quad(k+l>0),
\\
\Psi_{00}^\text{r}
&=-B_{00}\log t_1t_{-1}+o(\log t_1),
&\quad
\Psi_{kl}^\text{r}
&=B_{kl}t_1^{-k-l}+o(t_1^{-k-l})\quad(k+l>0).
}\label{Phikl-Psikl-t-small-leading}
$$
From the definition (\ref{Phi-Psi-func-def}) we obtain the recursion base for the coefficients:
\eq$$
A_{0l}={d A_l\over dt_0}={\tg\pi\nu\over\pi}{d A_l\over d\nu},
\qquad
A_{1l}=-\bar lA_l.
\qquad
B_{1l}=-\bar lB_l,
\label{A0l-A1l-explicit}
$$
Then by substituting (\ref{Phikl-Psikl-t-small-leading}) into the functional relations (\ref{Phikl-Psikl-recursion}) we obtain recursion relations for the coefficients:
\eq$$
\Aligned{
B_{kl}
&={2\nu\over k+l+1}A_{kl}+{\bar k\bar l\over(\overline{k+l})(k+l+1)}A_kA_l,
\\
A_{k+2,l}
&=(\overline{k+l})(k+l+1)\left(1-{4\nu^2\over(k+l+1)^2}\right)A_{kl}-2\bar k\bar l\left({1\over k+1}+{1\over k+l+1}\right)\nu A_kA_l.
}\label{BA-rec}
$$
Though we have not found an explicit closed formula for $A_{kl}$ with arbitrary $k$, $l$, these recursion relations enable one to find these values for any given $k$ and $l$. We have
\eq$$
W_{kl}(0)={K^\infty_{\nu,k}K^\infty_{\nu,l}\over8\lambda^2}{(-1)^{k+l-1}A_{kl}\over(k+l-1)!}
\qquad(k,l>0).
\label{Wkl-0-poskl}
$$
This proves (\ref{cVp-kl-fin}).

The situation with $\cI^-_{kl}$ is more complicated due to divergences as $t_0\to0$. Here we consider two cases: $l=1$ with arbitrary $k$ and $k=l=2$. In the case $l=1$ we may use the expansion (\ref{KIn-t-small}):
\Multline*$$
\Phi_{k,-1}=\d_{-1}\Phi_k
\\*
={2(-1)^k\,(k-1)!\,\lambda\over K^\infty_{\nu,k}}t_1^{-k}\,\d_{-1}\left(1+c^{(1)}_{\nu,-k}(4t_1t_{-1})^{1-2\nu}+c^{(1)}_{-\nu,-k}(4t_1t_{-1})^{1+2\nu}+O\left((t_1t_{-1})^{2-4|\nu|}\right)\right).
$$
Hence,
\eq$$
K_{\nu,k,-1}(t)=-{2^{k-2}\,(k-1)!\over\nu K^\infty_{\nu,k-1}}t^{1-k}\left((1-2\nu)c^{(1)}_{\nu,-k}t^{-4\nu}+(1+2\nu)c^{(1)}_{-\nu,-k}t^{4\nu}
+O\left(t^{2-8|\nu|}\right)\right).
\label{Knukm1-t-small}
$$
From this we derive
\eq$$
W_{k,-1}(t_0)=-{\beta_\nu^{-2}t_0^{-4\nu}+\beta_\nu^2t_0^{4\nu}\over8\nu}+O\left(t_0^{2-8|\nu|}\right).
\label{Wkm1-t0-small}
$$
This proves (\ref{cIm-k1-fin}).

In the case $k=l=2$ we need to use the small $t$ expansions up to $O(t^{6-12|\nu|})$. The small $t$ expansions for $\Phi_k^\text{r}$, $\Psi_k^\text{r}$ ($k=0,\pm1$), $\Phi_{00}^\text{r}$, $\Psi_{00}^\text{r}$ are obtained directly from the expansions for $\phi_\nu$, $\psi_\nu$, while those for $\Phi_{\pm2,0}^\text{r}$, $\Psi_{\pm2,0}^\text{r}$, $\Phi_{2,-2}^\text{r}$ are obtained by means of the relations (\ref{Phikl-Psikl-recursion}). We only succeeded to perform the calculation with the help of computer algebra, which resulted in~(\ref{cIm-22-fin}).

\section{Form factors in the semiclassical limit}
\label{sec:form-factors}

\subsection{General setting and simplest operators}
\label{sec:form-factors-simplest}

Consider $N$\-/point functions of the form
\eq$$
G_\cO(\{x_i\}_N)=\langle\varphi(x_N)\cdots\varphi(x_1)\cO(0)\rangle,
\label{G-gen-def}
$$
where $\cO(x)$ is any local operator in the sinh\-/Gordon model. Due to the standard reduction formulas it is possible to calculate the $n$\-/particle form factor:
\eq$$
F_\cO(\{\theta_i\}_N)
=\left(-\sqrt{4\pi Z_\varphi}\right)^{-N}\lim_{r_i\to\infty}\left(\prod^N_{i=1}r_i\int^{2\pi}_0d\xi_i\,\e^{\i mr_i\sh(\theta_i+\i\xi_i)}\lrd{r_i}\right)
G_\cO(\{x_i\}_N),
\label{FcO-GcO-rel}
$$
where $r_i,\xi_i$ are polar coordinates for $x_i$ on the Euclidean plane and the operator $\lrd t$ is defined as
$$
f(t)\lrd tg(t)={1\over2}(f(t)g'(t)-f'(t)g(t))
$$
and acts within the integrand. The factor $Z_\varphi$ is the wave function renormalization constant. In the leading order of the semiclassical expansion, one has $Z_\varphi=1$.

As a first example, let us consider the form factors of the operators $V_\nu$ defined above in~(\ref{Vnu-def}). In the leading order in $b$ we have
\Multline*$$
F_\nu(\{\theta_i\}_N)
=\left(-\sqrt{4\pi}\right)^{-N}\lim_{r_i\to\infty}\left(\prod^N_{i=1}r_i\int^{2\pi}_0d\xi_i\,\e^{\i mr_i\sh(\theta_i+\i\xi_i)}\lrd{r_i}\right)
\prod^N_{i=1}b^{-1}\phi_\nu(mr_i)
\\
=(-\sqrt2b^{-1}\lambda m)^N\lim_{r_i\to\infty}\prod^N_{i=1}
r_i\int^{2\pi}_0d\xi_i\,\left[\e^{\i t\sh(\theta_i+\i\xi_i)}\lrd tt^{-1/2}\e^{-t}\right]_{t=mr_i}
\\
=\left(\sqrt2b^{-1}\lambda\lim_{t\to\infty}\pi t(I_0(t)+I'_0(t))t^{-1/2}\e^{-t}\right)^N
=\left(\sqrt{4\pi}b^{-1}\lambda\right)^N.
$$
Here we used the trivial identity
$$
Px=-{m\over2}(\e^\theta z-\e^{-\theta}\bz)=-mr\sh(\theta+\i\xi)
$$
and the integral representation of the Bessel functions
\eq$$
I_k(t)=\int^{2\pi}_0{d\xi\over2\pi}\,e^{t\cos\xi}\cos k\xi.
\label{Ik-intrep}
$$
We considered all functions of the rapidities $\theta_i$ as analytic continuations from the imaginary axis.

Finally, we have
\eq$$
F_\nu(\{\theta_i\}_N)
=\left(\sqrt{4\pi}{\sin\pi\nu\over\pi b}\right)^N.
\label{ff-exp}
$$
In the limit $\nu=b\alpha\to0$, this gives $(\sqrt{4\pi}\alpha)^N$, which is consistent with the free field approximation. Moreover, it is consistent with the free field approximation in the vicinity of any point $\nu$. Indeed, let $|\delta|\ll1$. Then let us find the correlation function of the operator $\e^{b^{-1}\delta\chi}V_\nu$:
\Align*$$
\langle\infty|\T_r[\varphi(x_N)\ldots\varphi(x_1)]|b^{-1}\delta\rangle_\nu
&=\prod^N_{i=1}b^{-1}\phi_\nu(mr_i)+\sum^N_{i=1}\langle\infty|\chi(x_i)|b^{-1}\delta\rangle_\nu\,\prod_{j(\ne i)}b^{-1}\phi_\nu(mr_j)+O\left(\delta^2\right)
\\
&=\prod^N_{i=1}b^{-1}\phi_\nu(mr_i)+\sum^N_{i=1}4b^{-1}\delta K_{\nu,0}(mr_i)\,\prod_{j(\ne i)}b^{-1}\phi_\nu(mr_j)+O\left(\delta^2\right).
$$
Due to (\ref{K0-phi-rel}), the second term is the $\nu$ derivative of the first one. Hence
\eq$$
\langle\infty|\T_r[\varphi(x_N)\ldots\varphi(x_1)]|b^{-1}\delta\rangle_\nu
=\langle\infty|\T_r[\varphi(x_N)\ldots\varphi(x_1)]|0\rangle_{\nu+\delta}+O(\delta^2)
=G_{\nu+\delta}(\{x_i\}_N)+O(\delta^2).
\label{deltaQ-equivalence-large}
$$
We see that the correlation functions of $\e^{b^{-1}\delta\chi}V_\nu$ and $V_{\nu+\delta}$ with a number of operators $\varphi(x_i)$ coincide in the leading order in~$b$. The quantum oscillator contribution is equivalent to an appropriate modification of the classical solution. It extends the equivalence (\ref{deltaQ-equivalence-b2}) to larger shifts of the parameter $\nu$ for the special case $F[\varphi]=\prod_i\varphi(x_i)$.

Now let us study the operators of the form $\d^k\varphi\,V_\nu$, which correspond to the states $\i\,(k-1)!\,\a_{-k}|0\rangle_\nu$ in the radial picture. We have
\eq$$
\langle\infty|\T_r[\varphi(x_N)\ldots\varphi(x_1)]\a_{-k}|0\rangle_\nu
=\sum^N_{i=1}(-2\i)\e^{-\i k\xi_i}m^ku_k(mr_i)\prod_{j\ne i}b^{-1}\phi_\nu(mr_j)+\cdots.
\label{a-k-matel}
$$
To calculate the form factors we need to calculate the integrals over $\xi_i$. It can be done as follows:
\Multline$$
\int^{2\pi}_0{d\xi\over2\pi}\,\e^{\i t\sh(\theta+\i\xi)}\e^{-\i k\xi}
=\int^{2\pi}_0{d\xi\over2\pi}\,\e^{t\cos\xi}\e^{-k(\i\xi-\theta-{\i\pi\over2})}
=\i^k\e^{k\theta}\int^{2\pi}_0{d\xi\over2\pi}\,\e^{t\cos\xi}\e^{\i k\xi}
\\*
=\i^k\e^{k\theta}\int^{2\pi}_0{d\xi\over2\pi}\,\e^{t\cos\xi}\cos k\xi
=\i^k\e^{k\theta}I_k(t)\xrightarrow[t\to\infty]{}{\i^k\e^{k\theta}\over\sqrt{2\pi}}t^{-1/2}\e^t.
\label{int-to-Ik}
$$
Hence
\eq$$
F_{\d^k\varphi\,V_\nu}(\{\theta_i\}_N)
={b\over\nu}\left(\i m\over2\right)^k{K^\infty_{\nu,k}\over K^\infty_{\nu,1}}\sum^N_{i=1}\e^{k\theta_i}\>F_\nu(\{\theta_i\}_N).
\label{ffa-k}
$$
Note that the result for the operators $\bd^k\varphi\,V_\nu$ is obtained from this one by the substitution $m\to-m$, $\e^{k\theta_i}\to\e^{-k\theta_i}$. It is easy to see that the result (\ref{ffa-k}) in the simplest case $k=1$ is consistent with the identities
$$
\d\e^{\alpha\varphi}=\alpha\,\d\varphi\,\e^{\alpha\varphi},
\qquad
\bd\e^{\alpha\varphi}=\alpha\,\bd\varphi\,\e^{\alpha\varphi}.
$$

At first glance it may seem that the form factors of any descendant operator can be found as follows. The operators $\a_{-k},\ba_{-k}$ produce the following factors in the form factor:
\eq$$
\i\,(k-1)!\,\a_{-k}\to{b\over\nu}\left(\i m\over2\right)^k{K^\infty_{\nu,k}\over K^\infty_{\nu,1}}\e^{k\theta_i},
\qquad
\i\,(k-1)!\,\ba_{-k}\to{b\over\nu}\left(m\over2\i\right)^k{K^\infty_{\nu,k}\over K^\infty_{\nu,1}}\e^{-k\theta_i}.
\label{a-k->factor}
$$
The final summation should be taken over noncoinciding values of the $i$ subscripts. Nevertheless, we will see that this is not literally the case due to contributions of the interaction terms in the perturbation theory. It turns out that in form factors the interaction terms can produce contributions of the same order as the quantum fluctuations described by the quadratic part of the action. We will use the rule (\ref{a-k->factor}) in the final expressions that take into account the interaction terms.

In what follows, we will compare the results with the exact form factors~\cite{Feigin:2008hs,Lashkevich:2013mca,Lashkevich:2013yja}. The operators in this approach are enumerated by elements of a pair of commutative algebras $\cA$, $\bar\cA$ with the generators $c_{-k}$ and $\bc_{-k}$ ($k=1,2,\ldots$) correspondingly. The algebras are naturally graded by setting $\deg(c_{-k})=\deg(\bc_{-k})=k$. These two algebras are combined into one associative algebra $\cA^2$ with the commutation relation
\eq$$
[c_{-k},\bc_{-l}]={(1+(-1)^k)k\over4\sin^2{\pi pk\over2}}\delta_{kl},
\label{c-commut}
$$
where the parameter $p$ is defined in (\ref{m0-m-rel}). The operator $V^{h\bh'}_\nu(x)$ with $h\in\cA$, $\bh'\in\bar\cA$ is defined by a set of its form factors. It is a level $(\deg(h),\deg(h'))$ descendant of the operator $V^1_\nu=V_\nu$. Here we give the form factors of the operators $V^{h\bh'}_\nu$ in the limit $b\to0$ only:
\eq$$
f^{h\bh'}_\nu(\{\theta_i\}_N)=(\sqrt\pi b)^{-N}\sum_{\{\ve_i=\pm1\}_N}\e^{\i\pi({1\over2}-\nu)\sum^N_{i=1}\ve_i}
P^{h\bh'}_{\{\ve_i\}_N}(\{\e^{\theta_i}\}_N).
\label{f-hbh-def}
$$
Here the Laurent polynomials $P^g_{\{\ve_i\}_N}(\{x_i\}_N)$ ($g\in\cA^2$) are defined by the following set of relations:
\eq$$
\Gathered{
P^1_{\{\epsilon_i\}_N}(\{x_i\}_N)=1;
\\
P^{c_{-k}}_{\{\epsilon_i\}_N}(\{x_i\}_N)=\sum^N_{i=1}\ve_i^{k-1}x_i^k;
\qquad
P^{\bc_{-k}}_{\{\epsilon_i\}_N}(\{x_i\}_N)=\sum^N_{i=1}(-\ve_i)^{k-1}x_i^k;
\\
P^{k_1g_1+k_2g_2}_{\{\epsilon_i\}_N}(\{x_i\}_N)=k_1P^{g_1}_{\{\epsilon_i\}_N}(\{x_i\}_N)+k_2P^{g_2}_{\{\epsilon_i\}_N}(\{x_i\}_N),
\qquad
k_1,k_2\in\C,\ g_1,g_2\in\cA^2;
\\
P^{hh'}_{\{\epsilon_i\}_N}(\{x_i\}_N)=P^h_{\{\epsilon_i\}_N}(\{x_i\}_N)P^{h'}_{\{\epsilon_i\}_N}(\{x_i\}_N),
\qquad
h,h'\in\cA;
\\
P^{\bh\bh'}_{\{\epsilon_i\}_N}(\{x_i\}_N)=P^\bh_{\{\epsilon_i\}_N}(\{x_i\}_N)P^{\bh'}_{\{\epsilon_i\}_N}(\{x_i\}_N),
\qquad
\bh,\bh'\in\bar\cA;
\\
P^{\bh'h}_{\{\epsilon_i\}_N}(\{x_i\}_N)=P^h_{\{\epsilon_i\}_N}(\{x_i\}_N)P^{\bh'}_{\{\epsilon_i\}_N}(\{x_i\}_N),
\qquad
h\in\cA,\ \bh'\in\bar\cA.
}\label{Phbh-def}
$$
Let us stress that in the last line the order of $h$ and $\bh'$ is opposite to that in (\ref{f-hbh-def}). To exchange $h$ and $\bh'$ one has to use the commutation relation (\ref{c-commut}). This commutation leads to terms that cannot be reproduced directly in the semiclassical approximation. Nevertheless, later we will see a signature of the corresponding contributions in a simple example.

For $h=\bh'=1$ we have $P^{h\bh'}=1$ and the sum in the r.h.s.\ of (\ref{f-hbh-def}) reduces to $(2\sin\pi\nu)^N$. We thus recover the formula (\ref{ff-exp}) for the form factors $f^1_\nu$, which confirms the identification~\cite{Lukyanov:1997bp} of the operator $V^1_\nu$ with~$V_\nu$.

Note that the reflection properties in this limit are very simple. It is convenient to describe them in terms of an automorphism $r$ of $\cA^2$:
\eq$$
V^{h\bh'}_\nu=V^{r(h\bh')}_{1-\nu}=V^{r(h\bh')}_{-1-\nu},
\qquad
r(c_{-k})=(-1)^{k-1}c_{-k},
\qquad
r(\bc_{-k})=(-1)^{k-1}\bc_{-k}.
\label{Vhbh'-reflection}
$$

The action of $c_{-2k+1}$ and $\bc_{-2k+1}$ just multiplies any form factor by the sum $\sum^N_{i=1}\e^{\pm(2k-1)\theta_i}$. Hence
\eq$$
\Aligned{
f^{c_{-2k+1}}_\nu(\{\theta_i\}_N)
&=\sum^N_{i=1}\e^{(2k+1)\theta_i}F_\nu(\{\theta_i\}_N),
\\
f^{\bc_{-2k+1}}_\nu(\{\theta_i\}_N)
&=\sum^N_{i=1}\e^{-(2k+1)\theta_i}F_\nu(\{\theta_i\}_N).
}\label{f-c(2k-1)}
$$
Therefore,
\eq$$
\Aligned{
V^{c_{-2k+1}}_\nu
&=(-1)^k{\i\nu\over b}\left(2\over m\right)^{2k-1}{K^\infty_{\nu,1}\over K^\infty_{\nu,2k-1}}\,\d^{2k-1}\varphi\,V_\nu,
\\
V^{\bc_{-2k+1}}_\nu
&=(-1)^{k-1}{\i\nu\over b}\left(2\over m\right)^{2k-1}{K^\infty_{\nu,1}\over K^\infty_{\nu,2k-1}}\,\bd^{2k-1}\varphi\,V_\nu.
}\label{V-c(1-2k)}
$$
The element $c_{-2k}$ does not act on form factors directly. However, for $f^{c_{-2k}}_\nu$ it amounts to substituting one of the factors $2\sin\pi\nu$ by $2\i\cos\pi\nu$:
\eq$$
\Aligned{
f^{c_{-2k}}_\nu(\{\theta_i\}_N)
&=\i\ctg\pi\nu\,\sum^N_{i=1}\e^{2k\theta_i}F_\nu(\{\theta_i\}_N),
\\
f^{\bc_{-2k}}_\nu(\{\theta_i\}_N)
&=-\i\ctg\pi\nu\,\sum^N_{i=1}\e^{-2k\theta_i}F_\nu(\{\theta_i\}_N).
}\label{f-c(-2k)}
$$
Thus we have
\eq$$
\Aligned{
V^{c_{-2k}}_\nu
&=(-1)^k{\i\nu\over b\tg\pi\nu}\left(2\over m\right)^{2k}{K^\infty_{\nu,1}\over K^\infty_{\nu,2k}}\,\d^{2k}\varphi\,V_\nu,
\\
V^{\bc_{-2k}}_\nu
&=(-1)^{k-1}{\i\nu\over b\tg\pi\nu}\left(2\over m\right)^{2k}{K^\infty_{\nu,1}\over K^\infty_{\nu,2k}}\,\bd^{2k}\varphi\,V_\nu.
}\label{V-c(-2k)}
$$
It should be stressed that these relations and similar relations below are only correct in the leading order in~$b$. The exact relations must be more complicated and are generally still unknown.

We see that the operators $\d^k\varphi\,V_\nu$ (and $\bd^k\varphi\,V_\nu$) are not analytic under the reflections $\nu\to\pm1-\nu$. Nevertheless, we may interpret them as analytic continuations from the region $|\nu|<{1\over2}$. Then we obtain
\eq$$
\d^k\varphi\,V_\nu={K^\infty_{\nu,k}\over K^\infty_{\pm1-\nu,k}}\,\d^k\varphi\,V_{\pm1-\nu}.
\label{dkvarphi-V-reflection}
$$
This is evidently consistent with (\ref{Vhbh'-reflection}). In what follows we will establish relations between the operators $V^{h\bh'}_\nu$ and the operators obtained semiclassically. These relations will provide the easiest way for establishing the reflection properties from (\ref{Vhbh'-reflection}).

\subsection{Form factors of the operators \texorpdfstring{$\d^k\varphi\,\d^l\varphi\,V_\nu$}{dk varphi d varphi Vnu}}
\label{sec:dkphi-dlphi-V}

\begin{figure}[h]
\centering
\begin{subfigure}{.45\textwidth}
\centering
\begin{tikzpicture}[baseline,every node/.style={scale=.8}]
\tikzstyle{wave} = [decorate,decoration={snake,segment length=3pt,amplitude=.5pt}]
\draw[very thick] (0,0) -- (5,0) -- (5,1.2) -- (0,1.2) -- cycle;
\node at (2.5,.85) {$D_1\varphi\,D_2\varphi\,V_\nu$};
\node[rotate=90] (x1) at (.5,-2) {$b^{-1}\phi_\nu(x_1)$};
\draw (.5,0) -- (x1);
\node at (1.5,-1) {$\cdots$};
\node[rotate=90] (xNm2) at (2.5,-2) {$b^{-1}\phi_\nu(x_{N-2})$};
\draw (2.5,0) -- (xNm2);
\node[rotate=90] (xNm1) at (3.5,-2) {$\chi(x_{N-1})$};
\draw[wave] (3.5,0) node[above] {$D_1\varphi$} -- (xNm1);
\node[rotate=90] (xN) at (4.5,-2) {$\chi(x_N)$};
\draw[wave] (4.5,0) node[above] {$D_2\varphi$} -- (xN);
\end{tikzpicture}
\caption{}
\end{subfigure}
\begin{subfigure}{.45\textwidth}
\centering
\begin{tikzpicture}[baseline,every node/.style={scale=.8}]
\tikzstyle{wave} = [decorate,decoration={snake,segment length=3pt,amplitude=.5pt}]
\draw[very thick] (0,0) -- (5,0) -- (5,1.2) -- (0,1.2) -- cycle;
\node at (2.5,.85) {$D_1\varphi\,D_2\varphi\,V_\nu$};
\node[rotate=90] (x1) at (.5,-2) {$b^{-1}\phi_\nu(x_1)$};
\draw (.5,0) -- (x1);
\node at (1.5,-1) {$\cdots$};
\node[rotate=90] (xNm2) at (2.5,-2) {$b^{-1}\phi_\nu(x_{N-2})$};
\draw (2.5,0) -- (xNm2);
\node[rotate=90] (xNm1) at (3,-2) {$b^{-1}\phi_\nu(x_{N-1})$};
\draw (3,0) -- (xNm1);
\draw[fill] (4,-.7) coordinate (chi3) circle (2pt) node[right=2pt,yshift=-3pt] {$b\chi^3$};
\draw[wave] (3.5,0) node[above] {$D_1\varphi$} to[out=-90,in=150] (chi3);
\node[rotate=90] (xN) at (4,-2) {$\chi(x_N)$};
\draw[wave] (4.5,0) node[above] {$D_2\varphi$} to[out=-90,in=30] (chi3);
\draw[wave] (chi3) -- (xN);
\end{tikzpicture}
\caption{}
\end{subfigure}
\caption{The rectangles at the top denote the local operator $D_1\varphi\,D_2\varphi\,V_\nu$, where $D_i$ are any derivatives (e.g.\ $D_1=\d^k$, $D_2=\d^l$). The solid lines denote the classical solution $b^{-1}\phi_\nu$. Wavy lines denote the quantum pair correlation function $\langle\chi\chi\rangle$. Each diagram contains $N$ external lines of the correlation function. The symmetrization in the variables $x_i$ is assumed. In the diagram (a) all $N$ external lines are attached to the operator. $N-2$ lines correspond to the classical solution $b^{-1}\phi_\nu(x_i)$, while two other lines are paired to the quantum fields $\chi(x_i)$. In the diagram (b) $N-1$ external lines and $2$ internal ones are attached to the operator. The first $N-1$ external lines correspond to the classical solution, while the two others are quantum and linked to a three\-/particle vertex of order $b$, which connects them to the only external quantum line. The order of both diagrams is the same: $b^{2-N}$.}
\label{fig:twodiagrams}
\end{figure}
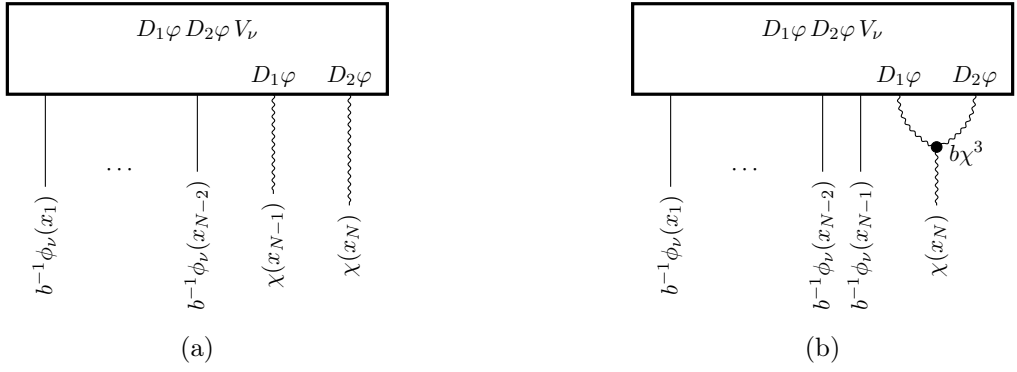

Now let us turn to the calculation of form factors of the operators that contain two fields in front of the exponential of the field. The operator $\d^k\varphi\,\d^l\varphi\,V_\nu$ corresponds to the state $-(k-1)!\,(l-1)!\,\a_{-k}\a_{-l}|0\rangle_\nu$ in the radial quantization picture. In the correlation functions, we should pair these two operators with two fields $\chi(x_i)$ thus providing an extra factor $b^2$ compared to the form factors of the exponential operator. However, this is just part of the story. We also need to take into account the interaction. Indeed, consider the two diagrams in Fig.~\ref{fig:twodiagrams}.
The first diagram (a) corresponds to the saddle point approximation, while the second diagram (b) contains one three\-/particle vertex due to the cubic term in~(\ref{shG-action-Delta}). Both diagrams are of the same order $b^{2-N}$ and thus must be taken into account in the leading order of correlation functions.

Then calculate the correlation function
\Align*$$
-{\textstyle{1\over(k-1)!\,(l-1)!}}
&G_{\d^k\varphi\,\d^l\varphi\,V_\nu}(\{x_i\}_N)
\\
&\qquad
=\langle\infty|\T_r\left[\varphi(x_N)\ldots\varphi(x_1)\left(1-S^{(3)}\right)\right]\a_{-k}\a_{-l}|0\rangle_\nu
\\
&\qquad
=\langle\infty|\T_r\left[\varphi(x_N)\ldots\varphi(x_1)
\left(1-{bm^2\over3!\cdot8\pi}\int d^2y\,\lcolon\chi^3(y)\rcolon\sh\phi_\mu(m|y|)\right)\right]\a_{-k}\a_{-l}|0\rangle_\nu
\\
&\qquad
=\langle\infty|\T_r[\varphi(x_N)\ldots\varphi(x_1)]\left(1+\i2^{-\delta_{kl}}{(k+l-1)!\over k!\,l!}b\cV^+_{kl}\a_{-k-l}\a_k\a_l\right)\a_{-k}\a_{-l}|0\rangle_\nu
\\
&\qquad
=\langle\infty|\T_r[\varphi(x_N)\ldots\varphi(x_1)]\a_{-k}\a_{-l}|0\rangle_\nu
\\*
&\qquad\quad
+4\i{(k+l-1)!\over(k-1)!\,(l-1)!}b\cV^+_{kl}\times\langle\infty|\T_r[\varphi(x_N)\ldots\varphi(x_1)]\a_{-k-l}|0\rangle_\nu.
$$
Here the interaction action $S^{(3)}$ is defined above in (\ref{S(n)-def}). Recall that the coefficients of the angular mode decomposition (\ref{chi-reddecomp}) of $\chi(x)$ are related to the functions $I_{\nu,k}(t)$, $K_{\nu,k}(t)$ by the equations (\ref{f-fact}), (\ref{u-KIn-final}), and $\cV^+_{kl}$ is defined by (\ref{cI-pm-def}), (\ref{cV-p-def}). We assume that the points $x_1,\ldots,x_N$ are far enough from the coordinate origin to take the $y$ integral on the whole plane without taking into account $r$ ordering with fields $\varphi(x_i)$. Besides, to get rid of unnecessary divergent loop terms we consider $\chi^n$ in the interaction terms as normal ordered products in the radial quantization sense.

By using the rules (\ref{a-k->factor}) and the expression (\ref{cVp-kl-fin}) for the integrals, we obtain
\eq$$
F_{\d^k\varphi\,\d^l\varphi\,V_\nu}(\{\theta_i\}_N)
={b^2\over\nu^2}\left(\i m\over2\right)^{k+l}{K^\infty_{\nu,k}K^\infty_{\nu,l}\over(K^\infty_{\nu,1})^2}
\left(\sum^N_{i,j=1}\e^{k\theta_i+l\theta_j}
-{K^\infty_{\nu,k+l}\over K^\infty_{\nu,kl}}\sum^N_{i=1}\e^{(k+l)\theta_i}\right)F_\nu(\{\theta_i\}_N).
\label{F-dkphi-dlphi}
$$
In particular, since $K^\infty_{\nu,k1}=K^\infty_{\nu,k}$, we find that
\eq$$
F_{\d^k\varphi\,\d\varphi\,V_\nu}(\{\theta_i\}_N)
={b^2\over\nu^2}\left(\i m\over2\right)^{k+1}\left({K^\infty_{\nu,k}\over K^\infty_{\nu,1}}\sum^N_{i,j=1}\e^{k\theta_i+\theta_j}
-{K^\infty_{\nu,k+1}\over K^\infty_{\nu,1}}\sum^N_{i=1}\e^{(k+1)\theta_i}\right)F_\nu(\{\theta_i\}_N).
\label{F-dkphi-dphi}
$$
This is consistent with the identity
$$
\d\left(\d^k\varphi\,\e^{\alpha\varphi}\right)=\d^{k+1}\varphi\,\e^{\alpha\varphi}+\alpha\,\d^k\varphi\,\d\varphi\,\e^{\alpha\varphi}.
$$

Now let us compare this result with the small $b$ limit of the exact form factors. We have
\eq$$
f^{c_{-k}c_{-l}}_\nu(\{\theta_i\}_N)=F_\nu(\{\theta_i\}_N)
\times\Cases{
\sum^N_{i,j=1}\e^{k\theta_i+l\theta_j},&\text{if $k,l\in2\Z+1$;}\\
\i\ctg\pi\nu\,\sum^N_{i,j=1}\e^{k\theta_i+l\theta_j},&\text{if $k-l\in2\Z+1$;}\\
\left(-\ctg^2\pi\nu\,\sum^N_{i\ne j}\e^{k\theta_i+l\theta_j}+\sum^N_{i=1}\e^{(k+l)\theta}\right),&\text{if $k,l\in2\Z$.}
}
\label{f-ckcl}
$$
From this we deduce the identification
\subeq{\label{V-ckcl}
\Align$$
V^{c_{-k}c_{-l}}_\nu
&=\left(2\over\i m\right)^{k+l}\left({\nu^2\over b^2}{(K^\infty_{\nu,1})^2\over K^\infty_{\nu,k}K^\infty_{\nu,l}}\,\d^k\varphi\,\d^l\varphi\,V_\nu
+{\nu\over b}{K^\infty_{\nu,1}\over K^\infty_{\nu,kl}}\,\d^{k+l}\varphi\,V_\nu\right),
&&\text{if $k,l\in2\Z+1$;}
\tag{\ref{V-ckcl}a}
\\
V^{c_{-k}c_{-l}}_\nu
&=\i\ctg\pi\nu\,\left(2\over\i m\right)^{k+l}\left({\nu^2\over b^2}{(K^\infty_{\nu,1})^2\over K^\infty_{\nu,k}K^\infty_{\nu,l}}\,\d^k\varphi\,\d^l\varphi\,V_\nu
+{\nu\over b}{K^\infty_{\nu,1}\over K^\infty_{\nu,kl}}\,\d^{k+l}\varphi\,V_\nu\right),
&&\text{if $k-l\in2\Z+1$;}
\tag{\ref{V-ckcl}b}
\\
V^{c_{-k}c_{-l}}_\nu
&\mathrlap{{}=\left(2\over\i m\right)^{k+l}\left(-{\nu^2\over b^2\tg^2\pi\nu}{(K^\infty_{\nu,1})^2\over K^\infty_{\nu,k}K^\infty_{\nu,l}}\,\d^k\varphi\,\d^l\varphi\,V_\nu
+{\nu K^\infty_{\nu,1}\over b\sin^2\pi\nu}\left({1\over K^\infty_{\nu,k+l}}
-{\cos^2\pi\nu\over K^\infty_{\nu,kl}}\right)\,\d^{k+l}\varphi\,V_\nu\right),}
\notag
\\
&\quad
&&\text{if $k,l\in2\Z$.}
\tag{\ref{V-ckcl}c}
$$
}
Here and below we do not write down separate equations for the opposite chirality. We will assume that they are obtained by the substitution $\d\leftrightarrow\bd$, $c_{-k}\leftrightarrow\bc_{-k}$, $b\to-b$.

\subsection{Renormalization and cancellation of the operators \texorpdfstring{$\d\bd\varphi\,V_\nu$}{d dbar Vnu}}
\label{sec:dbdphi}

Let us turn to the operators that contain descendants with two chiralities. In this subsection we will see that after an appropriate renormalization
\eq$$
\d\bd\varphi\,e^{\alpha\varphi}=0,
\quad
\text{if $\alpha\ne0$.}
\label{dbdphi-zero}
$$
Thus the following identity holds
\eq$$
\d\bd\e^{\alpha\varphi}=\alpha^2\,\d\varphi\,\bd\varphi\,\e^{\alpha\varphi}.
\label{dbd-exp-id}
$$

Define the `angular average' on the Euclidean plane over a small circle around the origin:
\eq$$
\overline{f(z,\bz)}_{r_0}=\int_{|z|=r_0}{dz\over2\pi\i z}\,f(z,\bz)=\int^{2\pi}_0{d\xi\over2\pi}\,f(r_0\e^{\i\xi},r_0\e^{-\i\xi}).
\label{angaverage-def}
$$
All operators without arguments will be assumed to be located at the origin.

Assume $0<|\nu|<1/2$, $mr_0\ll1$. Consider the product
$$
\overline{\upstrut\d\bd\varphi(z,\bz)}_{r_0}\,V_\nu
=\overline{\upstrut\d\bd\bigl(b^{-1}\phi_\nu(m|z|)+\chi(z,\bz)\bigr)}_{r_0}V_\nu
={m^2\over4}\left(b^{-1}\sh\phi_\nu(mr_0)+\overline{\chi(z,\bz)}_{r_0}\ch\phi_\nu(mr_0)\right)V_\nu,
$$
where we used the equations of motion for $\phi_\nu$ and $\chi$. In terms of the radial quantization, it reads
\Align*$$
\overline{\upstrut\d\bd\varphi(z,\bz)}_{r_0}|0\rangle_\nu
&={m^2\over4}\left(b^{-1}\sh\phi_\nu(mr_0)+\Q I_{\nu,0}(mr_0)\ch\phi_\nu(mr_0)\right)|0\rangle_\nu
\\
&={m^2\over8b}\left(\e^{\phi_\nu(mr_0)}(1+b\Q I_{\nu,0}(mr_0))-\e^{-\phi_\nu(mr_0)}(1-b\Q I_{\nu,0}(mr_0))\right)|0\rangle_\nu
\\
&={m^2\over8b}\left(\e^{\phi_\nu(mr_0)}|0\rangle_{\nu+b^2}-\e^{-\phi_\nu(mr_0)}|0\rangle_{\nu-b^2}\right).
$$
In the last line, we used the approximations $I_{\nu,0}(t)=1+O\left(t^{2-4|\nu|}\right)$, $1\pm b\Q\simeq\e^{\pm b\Q}$ and the equivalence (\ref{deltaQ-equivalence-b2}). Since $\e^{-\phi_\nu(t)}=\beta_\nu^2t^{4\nu}(1+O(t^{2-4\nu}))$, the second term in the parentheses vanishes as $r_0\to0$ for $\nu>0$. The first term, on the contrary, diverges as $r_0^{-4\nu}$. These properties are flipped for $\nu<0$.

Let us recall the key facts established in the framework of the conformal perturbation theory\cite{Zamolodchikov:1990bk}. Its starting point is a conformal field theory. The perturbation operator $\Phi_\text{p}$ is a primary spinless operator of the conformal dimension $\Delta_\text{p}<1$. Consider any operator $\cO(x)$ with given conformal dimensions $(\Delta_\cO,\bar\Delta_\cO)$. Suppose that there are operators $\cO^{(n)}_i(x)$ ($n>0$) of conformal dimensions $(\Delta^{(n)}_i,\bar\Delta^{(n)}_i)$ that appear in the multiple operator product expansion $\cO(x)\Phi_\text{p}(y_1)\cdots\Phi_\text{p}(y_n)$, so that
\eq$$
\delta^{(n)}_i=\Delta_\cO-\Delta^{(n)}_i-n(1-\Delta_\text{p})=\bar\Delta_\cO-\bar\Delta^{(n)}_i-n(1-\Delta_\text{p})\ge0.
\label{AlZ-inequality}
$$
Then the operator $\cO(x)$ should be renormalized by adding the operators $\cO^{(n)}_i$:
\eq$$
\cO_\text{ren}(x)=\cO(x)-\sum_{n>0}\sum_i U^{(n)}_ir_0^{-2\delta^{(n)}_i}\cO^{(n)}_i(x),
\label{cO-renorm}
$$
where $r_0$ is a small distance corresponding to the ultraviolet cutoff and $U^{(n)}_i$ are certain constants. If the inequality in (\ref{AlZ-inequality}) turns into an equality for some particular $n$ and $i$, the corresponding term in (\ref{cO-renorm}) should be substituted by
$$
U^{(n)}_i\to U^{(n)}_i\log{r_0\over\rho},
$$
where $\rho$ is an additional dimensional parameter that expresses nonuniqueness of the renormalization. This situation is called an operator resonance.

Now return to the sinh\-/Gordon model. It can be considered as a conformal free massless boson theory perturbed by the operator $\Phi_\text{p}=\ch b\varphi$ with the conformal dimension $\Delta_\text{p}=-b^2$. Since we consider small values of $b$, the small coupling region in the sense of the conformal perturbation theory corresponds to the scales $r\ll bm^{-1}$. But the semiclassical description deals with the scales $r\gg bm^{-1}$, and the ultraviolet cutoff $r_0$ lies in this region. Thus, though the structure (\ref{cO-renorm}) is preserved in the semiclassical limit, the coefficients $U^{(n)}_i$ may not coincide with those of the conformal perturbation theory. Moreover, they may depend on $r_0$, but must remain finite in the formal limit $r_0\to0$.

Consider the operator  $\cO=\d\bd\varphi\,\e^{\alpha\varphi}$ ($\alpha=b^{-1}\nu$) with the conformal dimensions $\Delta_\cO=\bar\Delta_\cO=1-\alpha^2$. For $n=1$ there are two operators $\cO^{(1)}_1=\e^{(\alpha+b)\varphi}$ and $\cO^{(1)}_2=\e^{(\alpha-b)\varphi}$ that appear in the operator product expansion $\cO(x)\Phi_\text{p}(y)$ and have necessary dimensions for some values of $\nu$. We have $\delta^{(1)}_1=4\nu$, $\delta^{(1)}_2=-4\nu$. Thus the operator $\cO^{(1)}_1$ must be added if $\nu>0$, while the operator $\cO^{(1)}_2$ must be added if $\nu<0$. Let us add both of them to define the renormalized operator independently of the sign of $\nu$. To get rid of some subleading divergences that appear for $|\nu|>1/4$ we will add operators with the coefficients $\e^{\pm\phi_\nu}$ instead of $r_0^{\mp4\nu}$. Thus let us define the renormalized operator as follows:
\Multline$$
[\d\bd\varphi\,V_\nu]_\text{ren}
=\left.\left(\overline{\upstrut\d\bd\varphi(z,\bz)}_{r_0}V_\nu
  -{m^2\over8b}\e^{\phi_\nu(mr_0)}V_{\nu+b^2}-{m^2\over8b}\e^{-\phi_\nu(mr_0)}V_{\nu-b^2}
\right)\right|_{r_0\to0}
\\
=\left.\left(\overline{\upstrut\d\bd\varphi(z,\bz)}_{r_0}
  -{m^2\over8b}\e^{\phi_\nu(mr_0)}\left(1+b\overline{\chi(z,\bz)}_{r_0}\right)
  -{m^2\over8b}\e^{-\phi_\nu(mr_0)}\left(1-b\overline{\chi(z,\bz)}_{r_0}\right)
\right)V_\nu\right|_{r_0\to0}=0.
\label{dbdphi-reg-def}
$$
This proves~(\ref{dbdphi-zero}).

The case $\nu=0$ is exceptional. In this case both coefficients at $V_{b^2}$ and $V_{-b^2}$ are finite, and the operator $\d\bd\varphi\,V_0=\d\bd\varphi$ does not need renormalization. We return to the equation of motion $\d\bd\varphi={m^2\over8b}\sin b\varphi$.

\subsection{Form factors of the operators \texorpdfstring{$\d\varphi\,\bd\varphi\,V_\nu$}{d varphi dbar varphi Vnu}}
\label{sec:dphi-bdphi-V}

Let us begin with a more accurate derivation of the radial quantization description of the operator $\d\varphi\,V_\nu$. Define
$$
\d\varphi\,V_\nu=\lim_{r_0\to0}\overline{\d\varphi(z,\bz)}_{r_0}V_\nu
=\lim_{r_0\to0}\overline{(b^{-1}\,\d\phi_\nu(m|z|)+\d\chi(z,\bz))}_{r_0}V_\nu
=\lim_{r_0\to0}\overline{\d\chi(z,\bz)}_{r_0}V_\nu.
$$
The derivative of $\phi_\nu$ vanishes after the angular averaging, while in $\d\chi$ only one term $\d(\i \a_{-1}f_{-1}(z,\bz))=\i \a_{-1}+O\left(r_0^{2-4|\nu|}\right)$ survives. Hence, this operator is given by $\i \a_{-1}|0\rangle_\nu$ in the radial quantization, as it was postulated while deriving (\ref{ffa-k}).

Now consider the operator $\d\varphi\,\bd\varphi\,V_\nu$. It should be extracted from the operator
$$
\overline{\upstrut\bd\varphi(z_2,\bz_2)}_{r_{02}}\overline{\d\varphi(z_1,\bz_1)}_{r_{01}}V_\nu.
$$
Assume, for example, that $r_{01}<r_{02}$. Due to the radial ordering it means that the operator $\bd\varphi$ is placed to the left of $\d\varphi$ in the radial operator product. Then, as we have just seen, we may put $r_{01}=0$ and substitute $\d\varphi\,V_\nu$ by $\i\a_{-1}|0\rangle_\nu$. Thus we should calculate $\overline{\upstrut\bd\varphi(z,\bz)}_{r_0}\overline{\upstrut\d\varphi(z_1,\bz_1)}_0|0\rangle_\nu=\overline{\upstrut\bd\varphi(z,\bz)}_{r_0}\i \a_{-1}|0\rangle_\nu$ with $r_0=r_{02}$. But the rotationally invariant part of the operator $\bd\varphi(z,\bz)$ is the combination $\i\ba_{-1}\bd\bar f_{-1}(z,\bz)-\i \a_1\bd f_1(z,\bz)$. From (\ref{KIn-t-small}) we have
$$
\Aligned{
\bar f_{-1}(z,\bz)
&=\bz+O\left(r^{2-4\nu}\right),
\\
f_1(z,\bz)
&=z^{-1}-{\beta_\nu^{-2}m^{2-4\nu}\over16\nu(1-2\nu)}z^{-2\nu}\bz^{1-2\nu}-{\beta_\nu^2m^{2+4\nu}\over16\nu(1+2\nu)}z^{2\nu}\bz^{1+2\nu}+\bz o(|z|^{-4|\nu|}).
}
$$
Since $\a_1\a_{-1}|0\rangle_\nu=2|0\rangle_\nu$, we obtain
$$
\overline{\upstrut\bd\varphi(z,\bz)}_{r_0}\overline{\upstrut\d\varphi(z_1,\bz_1)}_0|0\rangle_\nu
=\left(-\a_{-1}\ba_{-1}-{\beta_\nu^{-2}m^{2-4\nu}\over8\nu}r_0^{-4\nu}+{\beta_\nu^2m^{2+4\nu}\over8\nu}r_0^{4\nu}\right)|0\rangle_\nu.
$$
Depending on the sign of $\nu$, one of the last two terms is divergent as $r_0\to0$. To cancel this divergence, we have to add the operators $V_{\nu\pm b^2}$ with appropriate coefficients:
\eq$$
\left[\bd\varphi\,\d\varphi\,V_\nu\right]_\text{ren}
=\left.\overline{\upstrut\bd\varphi(z,\bz)}_{r_0}\d\varphi\,V_\nu
+{\beta_\nu^{-2}m^2\over8\nu}(mr_0)^{-4\nu}V_{\nu+b^2}-{\beta_\nu^2m^2\over8\nu}(mr_0)^{4\nu}V_{\nu-b^2}\right|_{r_0\to0}.
\label{dphi-bdphi-V-ren}
$$
Then, expanding in $b$, we obtain
\Multline$$
\left[\overline{\upstrut\bd\varphi(z,\bz)}_{r_0}\overline{\upstrut\d\varphi(z_1,\bz_1)}_0|0\rangle_\nu\right]_\text{ren}
\\
=\overline{\upstrut\bd\varphi(z,\bz)}_{r_0}\overline{\upstrut\d\varphi(z_1,\bz_1)}_0|0\rangle_\nu
+{\beta_\nu^{-2}m^2\over8\nu}(mr_0)^{-4\nu}(1+b\Q)|0\rangle_\nu-{\beta_\nu^2m^2\over8\nu}(mr_0)^{4\nu}(1-b\Q)|0\rangle_\nu
\\
=\left(-\a_{-1}\ba_{-1}+{bm^2\over8\nu}\left(\beta_\nu^{-2}(mr_0)^{-4\nu}+\beta_\nu^2(mr_0)^{4\nu}\right)\Q\right)|0\rangle_\nu.
\label{dphi-bdphi-V-ren-zerothorder}
$$
Being of the first order in $b$, the second term in the last line seems to be negligible, but this is not the case. Indeed, the form factors of the first term are of order $b^2$ compared to those of $V_\nu$ due to the two factors $\langle\chi(x_i)\a_{-1}\rangle$ and $\langle\chi(x_j)\ba_{-1}\rangle$ instead of $b^{-1}\phi_\nu(mx_i)$ and $b^{-1}\phi_\nu(mx_j)$. The second term substitutes only one factor $b^{-1}\phi_\nu(mx_i)$ by a factor $\sim b\langle\chi(x_i)\Q\rangle$. Therefore in the form factors both terms are \emph{of the same order}. Besides, the second term is divergent.

\textbf{Remark.} In fact, the corrections to this expression are of the order $O\left(r_0^{2-8|\nu|}\right)$. It means that for ${1\over4}\le|\nu|<{1\over2}$ more divergent terms appear. In what follows we will, in fact, assume that $|\nu|<{1\over4}$ and continue the finite part of the results to the whole region $|\nu|<{1\over2}$ analytically.

The divergent term is canceled when we take into account the diagram in Fig.~\ref{fig:twodiagrams}b with a three\-/leg vertex. The calculation of this contribution demands some accuracy. It is important that first we need to calculate the corrections for the product $\varphi(z_2,\bz_2)\varphi(z_1,\bz_1)$ with $|z_2|>|z_1|>0$ and only after that to take the spatial derivatives.

This procedure may be simplified using the angular averaging. We have
\Multline$$
\T_r[-S^{(3)}\overline{\upstrut\bd\varphi(z_2,\bz_2)}_{r_{02}}\overline{\upstrut\d\varphi(z_1,\bz_1)}_{r_{01}}]|0\rangle_\nu
\\*
=-\bd_2\d_1\T_r\left[S^{(3)}
(\i\ba_{-1}\bar f_{-1}(z_2,\bz_2)-\i\a_1f_1(z_2,\bz_2))\i\a_{-1}f_{-1}(z_1,\bz_1)\right]|0\rangle_\nu\biggr|_{z_i=r_{0i}}.
\label{dphi-bdphi-V-firstorder}
$$
We deliberately bring the interaction action $S^{(3)}\sim\int d^2y\,\chi^3(y)\sh\phi_\nu(m|y|)$ inside the derivatives, because the radial ordering $\T_r$ puts $\chi^3(y)$ in different positions in the regions $|y|<|z_1|$, $|z_1|<|y|<|z_2|$, $|z_2|<|y|$, which contributes the $z_i$ and $\bz_i$ dependence of the result. It is easy to check that the contribution of the region $|y|<|z_1|$ is vanishing, so that we may take the $\d_1$ derivative and then set $|z_1|=0$:
$$
{bm^2\over3!\cdot8\pi}\,\bd\T_r\left[\int d^2y\,\chi^3(y)\sh\phi_\nu(m|y|)\,(\ba_{-1}\bar f_{-1}(z,\bz)-\a_1f_1(z,\bz))\right]\a_{-1}|0\rangle_\nu.
$$
Here we assumed that $z=z_2$. Taking the leading order in $f_k$ and rewriting the $r$ ordering explicitly, we obtain
\Multline*$$
{bm^2\over3!\cdot8\pi}\,\bd\left(\int_{|y|>|z|}d^2y\,\chi^3(y)\bz\ba_{-1}\a_{-1}\sh\phi_\nu(m|y|)
-\int_{|y|<|z|}d^2y\,z^{-1}\a_1\chi^3(y)\a_{-1}\sh\phi_\nu(m|y|)\right)|0\rangle_\nu
\\
=-{bm^2\over8\pi}\,\bd\biggl(\bz\int_{|y|>|z|}d^2y\,\Q\a_1f_1(y)\ba_1\bar f_1(y)\ba_{-1}\a_{-1}\sh\phi_\nu(m|y|)
\\*
+z^{-1}\int_{|y|<|z|}d^2y\,\a_1\a_{-1}f_{-1}(y)\Q\a_1f_1(y)\a_{-1}\sh\phi_\nu(m|y|)\biggr)|0\rangle_\nu
\\
=-b\,\bd\left(m^2\bz\cI^-_{11}(m|z|)+2z^{-1}\cI^0_{11}(m|z|)\right)\Q|0\rangle_\nu.
$$
In the first term, the operator $\chi^3$ reduces to a single contribution $-3!\,\Q\a_1f_1\ba_1\bar f_1$, while in the second term it yields $3!\,\a_{-1}f_{-1}\Q\a_1f_1$. Combinations of the operators $\a_{\pm1},\ba_{\pm1}$ reduce to numbers in both terms.

Now we can use the explicit expressions (\ref{cI-0-explicit}), (\ref{cIm-k1-fin}). Collecting all terms and taking the $\bd$ derivative, we find that
$$
\T_r[-S^{(3)}\overline{\upstrut\bd\varphi(z,\bz)}_{r_0}\overline{\upstrut\d\varphi(z_1,z_1)}_0]|0\rangle_\nu
=\left(bm^2{\tg\pi\nu\over4\nu^2}+{bm^2\over8\nu}\left(\beta_\nu^{-2}(mr_0)^{-4\nu}+\beta_\nu^2(mr_0)^{4\nu}\right)\right)\Q|0\rangle_\nu,
$$
where $r_0=|z|$. The second term in the parentheses exactly cancels the divergent part in (\ref{dphi-bdphi-V-ren-zerothorder}). Thus we have
\eq$$
\left[\T_r[\e^{-S_\text{int}[\chi]}\overline{\upstrut\bd\varphi(z,\bz)}_{r_0}\overline{\upstrut\d\varphi(z_1,z_1)}_0]|0\rangle_\nu\right]_\text{ren}
=\left(-\a_{-1}\ba_{-1}+bm^2{\tg\pi\nu\over4\pi\nu^2}\Q\right)|0\rangle_\nu.
\label{dphi-bdphi-reg}
$$
For the correlation function we obtain
\Multline$$
\langle\infty|\T_r[\varphi(x_N)\cdots\varphi(x_1)]\left(-\a_{-1}\ba_{-1}+bm^2{\tg\pi\nu\over4\pi\nu^2}\Q\right)|0\rangle_\nu
\\
=-\sum_{1\le i\ne j\le N}(-2\i)\e^{-\i\xi_i}mu_1(mr_i)(-2\i)\e^{\i\xi_j}mu_1(mr_j)\prod_{k\ne i,j}b^{-1}\phi_\nu(mr_k)
\\*
+bm^2{\tg\pi\nu\over4\pi\nu^2}\sum^N_{i=1}4u_*(mr_i)\prod_{k\ne i}b^{-1}\phi_\nu(mr_k).
\label{dphi-bdphi-corr}
$$
After applying the rule (\ref{FcO-GcO-rel}) we see that the contribution of the second term just completes the sum over $i,j$ by the terms with $i=j$:
\eq$$
F_{\d\varphi\,\bd\varphi\,V_\nu}(\{\theta_i\}_N)
=\left(bm\over2\nu\right)^2\sum^N_{i,j=1}\e^{\theta_i-\theta_j}\>F_\nu(\{\theta_i\}_N)
\label{dphi-bdphi-ff}
$$
in consistency with~(\ref{dbd-exp-id}). In other words,
\eq$$
V^{c_{-1}\bar c_{-1}}_\nu=\left(2\nu\over bm\right)^2\d\varphi\,\bd\varphi\,V_\nu,
\label{V-c1bc1}
$$
as it was expected.

\subsection{Form factors of the operators \texorpdfstring{$\d^k\varphi\,\bd\varphi\,V_\nu$}{dk varphi dbar varphi Vnu}}
\label{sec:dkphi-bardphi-V}

Now let us calculate the form factors of the operators $\d^k\varphi\,\bd\varphi\,V_\nu$ for $k>1$. Since $\a_1\a_{-k}|0\rangle_\nu=0$ in this case, these operators do not demand renormalization in the zeroth order of the perturbation theory. However, as we will see below, they require renormalization in the first order.

In the radial quantization, the operator $\d^k\varphi\,\bd\varphi\,V_\nu$ corresponds to the state $-(k-1)!\,\a_{-k}\ba_{-1}|0\rangle_\nu$ in the zeroth order. As for the previous case, the first two orders can be found by calculating $\overline{\upstrut\bd\varphi(z,\bz)}_{r_0}\overline{\upstrut\d^k\varphi(z_1,\bz_1)}_0|0\rangle_\nu$. Thus we obtain
\Align*$$
&-{\textstyle{1\over(k-1)!}}G_{\d^k\varphi\,\bd\varphi\,V_\nu}(\{x_i\}_N)
\\
&\qquad
=\bd\langle\infty|\T_r\left[\varphi(x_N)\ldots\varphi(x_1)\left(1-S^{(3)}\right)
(\ba_{-1}\bar f_{-1}(z,\bz)-\a_1f_1(z,\bz))\right]\a_{-k}|0\rangle_\nu
\\
&\qquad
=\langle\infty|\T_r[\varphi(x_N)\ldots\varphi(x_1)]
\\*
&\qquad\quad
\times\bd\left(\bz\ba_{-1}
+{\i bm^2\bz\over4k(k-1)}\cI^-_{k1}(m|z|)\a_{-k+1}\a_k\ba_1\ba_{-1}+{\i bz^{-1}\over2k(k-1)}\cI^0_{k1}(m|z|)\a_1\a_{-k+1}\a_k\a_{-1}\right)\a_{-k}|0\rangle_\nu
\\
&\qquad
=\langle\infty|\T_r[\varphi(x_N)\ldots\varphi(x_1)]\left(\a_{-k}\ba_{-1}
+{\i bm^2\over k-1}\left(\cV^-_{k1}-{\beta_\nu^{-2}(mr_0)^{-4\nu}+\beta_\nu^2(mr_0)^{4\nu}\over8\nu}\right)\a_{-k+1}\right)|0\rangle_\nu,
$$
where we assume that $|z|=r_0$. Due to the presence of a divergent term, we need to renormalize the operator by adding the operators $\d^{k-1}\varphi\,V_{\nu\pm b^2}$ with appropriate coefficients:
\Multline$$
\left[\d^k\varphi(z,\bz)\,\bd\varphi\,V_\nu\right]_{\rm ren}
=\overline{\upstrut\bd\varphi(z,\bz)}_{r_0}\d^k\varphi\,V_\nu
\\*
-{b\over8\nu}\left(\beta_\nu^{-2}(mr_0)^{-4\nu}\,\d^{k-1}\varphi\,V_{\nu+b^2}+\beta_\nu^2(mr_0)^{4\nu}\,\d^{k-1}\varphi\,V_{\nu-b^2}\right)\biggr|_{r_0\to0}.
\label{dkphi-bdphi-V-ren}
$$
In the r.h.s.\ we may substitute $V_{\nu\pm b^2}$ by $V_\nu$ since the difference is of order $b^2$ compared to the other terms. The renormalization simply removes the divergent term in the correlation function, and we have
$$
G^{\rm ren}_{\d^k\varphi\,\bd\varphi\,V_\nu}(\{x_i\}_N)
=\langle\infty|\varphi(x_N)\ldots\varphi(x_1)\left(\i(k-1)!\,\a_{-k}\,\i\ba_{-1}
+{bm^2\over4\nu}{K^\infty_{\nu,k}\over K^\infty_{\nu,k-1}}\i(k-2)!\,\a_{-k+1}\right)|0\rangle_\nu.
$$
Calculating the large\-/distance asymptotics, we obtain
\eq$$
F_{\d^k\varphi\,\bd\varphi\,V_\nu}(\{\theta_i\}_N)
=-{b^2\over\nu^2}\left(\i m\over2\right)^{k+1}{K^\infty_{\nu,k}\over K^\infty_{\nu,1}}
\sum^N_{i,j=1}\e^{k\theta_i-\theta_j}\>F_\nu(\{\theta_i\}_N).
\label{F-dkphi-bdphi}
$$
This means that
\eq$$
\bd\left(\d^k\varphi\,\e^{\alpha\varphi}\right)=\alpha\,\d^k\varphi\,\bd\varphi\,\e^{\alpha\varphi}
\quad\Leftrightarrow\quad
\d^k\bd\varphi\,\e^{\alpha\varphi}=0
\label{dkbd-exp-id}
$$
for $\alpha\ne0$. We may conjecture that any descendant of a non\-/identity operator that contains $\d\bd\varphi$ vanishes.

\subsection{Operators \texorpdfstring{$\d^k\varphi\,\bd^k\varphi\,V_\nu$}{dk varphi dbark varphi Vnu}: cancellation of the leading divergence}
\label{sec:dkphi-bdkphi-V}

Consider the operators $\d^k\varphi\,\bd^k\varphi\,V_\nu$ for arbitrary values of $k$. In the zeroth order in the perturbation theory we have
\eq$$
\overline{\upstrut\bd^k\varphi(z,\bz)}_{r_0}\overline{\upstrut\d^k\varphi(z_1,\bz_1)}_0|0\rangle_\nu
=-{(k-1)!\over k}\bd^k(\ba_{-k}\bar f_{-k}(z,\bz)-\a_kf_k(z,\bz))\a_{-k}|0\rangle_\nu
\label{dkphi-bdkphi-V-zerothorder}
$$
Now calculate the first order contribution:
\Multline$$
\T_r\left[-S^{(3)}\overline{\upstrut\bd^k\varphi(z,\bz)}_{r_0}\,\overline{\upstrut\d^k\varphi(z_1,\bz_1)}_0\right]|0\rangle_\nu
={(k-1)!\over k}\,\bd^k\T_r\left[S^{(3)}
(\ba_{-k}\bar f_{-k}(z,\bz)-\a_kf_k(z,\bz))\a_{-k}\right]|0\rangle_\nu
\\
=-{(k-1)!\over k}bm^2\,\bd^k\left({2^{2-2k}\over(k-1)!^2}m^{2k}\bar f_{-k}(z,\bz)\cI^-_{kk}(m|z|)+2kf_k(z,\bz)\cI^0_{kk}(m|z|)\right)\Q|0\rangle_\nu.
\label{dkphi-bdkphi-V-firstorder}
$$
The calculation of arbitrary integrals $\cI^-_{kk}$ is not an easy task and has not been made in general case. Here we only discuss the leading singularities and in the next subsection we will fully consider the case $k=2$. For $\bar f_{-k}$ we only take into account the leading singularity $\bz^k$, while for $f_k$ the subleading contributions of the order $z^{-k}|z|^{2\pm4\nu}$ contribute to the zeroth order of the perturbation theory. We obtain
\eq$$
\Gathered{
\overline{\upstrut\bd^k\varphi(z,\bz)}_{r_0}\overline{\upstrut\d^k\varphi(z_1,\bz_1)}_0|0\rangle_\nu
=\left(\tilde c_{\nu,k}r_0^{-2k+2-4\nu}+\tilde c_{-\nu,k}r_0^{-2k+2+4\nu}+O(r_0^{-2k+4-8|\nu|})\right)|0\rangle_\nu,
\\
\tilde c_{\nu,k}=-{(k-1)!\,\beta_\nu^{-2}m^{2-4\nu}\over4(k-1+2\nu)}{\Gamma(1-2\nu)\over\Gamma(2-k-2\nu)}.
}\label{dkphi-bdkphi-V-zerothorder-leading}
$$
The leading contribution in $r_0$ to the first order term (\ref{dkphi-bdkphi-V-firstorder}) comes from the leading contributions of $\bar f_{-k}$, $f_k$, $\cI^-_{kk}$ and $\cI^0_{kk}$. A straightforward calculation gives
\eq$$
\T_r\left[-S^{(3)}\overline{\upstrut\bd^k\varphi(z,\bz)}_{r_0}\overline{\upstrut\d^k\varphi(z_1,\bz_1)}_0\right]|0\rangle_\nu
=\left(\tilde c_{\nu,k}r_0^{-2k+2-4\nu}-\tilde c_{-\nu,k}r_0^{-2k+2+4\nu}+O(r_0^{-2k+4-8|\nu|})\right)\Q|0\rangle_\nu.
\label{dkphi-bdkphi-V-firstorder-leading}
$$

Note that all divergent parts come in the form of a linear combination of $|0\rangle_\nu$ and $\Q|0\rangle_\nu$. It means that in the leading order of the semiclassical perturbation theory the counterterms must be of the form $V_{\nu+sb^2}$ with  $s\sim1$. From the conformal perturbation theory we know that $s$ must be an integer, $|s|$ must be not greater than the order of the conformal perturbation theory $n$, and $n-s$ must be even. Thus we conjecture that
\eq$$
\left[\d^k\varphi\,\bd^k\varphi\,V_\nu\right]_\text{ren}
=\overline{\upstrut\bd^k\varphi(z,\bz)}_{r_0}\overline{\upstrut\d^k\varphi(z_1,\bz_1)}_0\,V_\nu
-\sum^k_{n=1}\sum^n_{\substack{s=-n\\n-s\in2\Z}}F^{(n,s)}_{\nu,k}(mr_0)r_0^{2n-2k-4s\nu}V_{\nu+sb^2}\biggr|_{r_0\to0}
\label{dkphi-bdkphi-V-ren}
$$
with certain coefficients $F^{(n,s)}_{\nu,k}(t)$, which are polynomials in the variables $t$ and $\log t$ considered independently. From equations (\ref{dkphi-bdkphi-V-zerothorder-leading}), (\ref{dkphi-bdkphi-V-firstorder-leading}) we conclude that
\eq$$
F^{(1,\pm1)}_{\nu,k}(0)=\tilde c_{\pm\nu,k}.
\label{tildec-1pm1}
$$
Most of the coefficients $F^{(n,s)}_{\nu,k}(t)$ are finite at $t=0$ with one exception. The power of $r_0$ corresponding to $n=k$, $s=0$ is formally zero. However, from the conformal perturbation theory we know that it is $2kb^2$. Formally $r_0^{2kb^2}$ is not divergent. It becomes divergent in the sine\-/Gordon region $b^2<0$. It means that for small $b$ it is very close to a resonance. This results into a logarithmic term in the semiclassical calculation:
\eq$$
F^{(k,0)_{\nu,k}}(t)=\tilde c'_{\nu,k}\log t+O(1)
\label{F(k,0)-log}
$$
with a certain coefficient $\tilde c'_{\nu,k}$. Due to the proximity to a resonance, this logarithm should be treated as a limit of a difference:
\eq$$
\log mr_0\simeq{(mr_0)^{2kb^2}-1\over2kb^2}.
\label{log-to-power}
$$
If, instead of subtracting $F^{(k,0)}_{\nu,k}(mr_0)$, we subtract only $\tilde c'_{\nu,k}(mr_0)^{2kb^2}/2kb^2$, the renormalized operator will change into
\eq$$
\left[\d^k\varphi\,\bd^k\varphi\,V_\nu\right]'_\text{ren}=\left[\d^k\varphi\,\bd^k\varphi\,V_\nu\right]_\text{ren}+{\tilde c'_{\nu,k}\over2kb^2}V_\nu.
\label{dkphi-bdkphi-vac-contrib}
$$
This will provide a vacuum expectation value of the renormalized descendant operator, which is qualitatively consistent with the known results\cite{Fateev:1998xb,Jimbo:2009ja,Negro:2013wga}. We shall see that in the case $k=2$ it is consistent quantitatively. The contribution (\ref{dkphi-bdkphi-vac-contrib}) is of order $b^{-4}$ comparing to all other terms, so that to fully describe this contribution we would have to find further perturbative corrections.

\subsection{Form factors of the operators \texorpdfstring{$\d^2\varphi\,\bd^2\varphi\,V_\nu$}{d2 varphi dbar2 varphi Vnu}}
\label{sec:d2phi-bd2phi-V}

Finally, let us now calculate the right hand sides of equations (\ref{dkphi-bdkphi-V-zerothorder}), (\ref{dkphi-bdkphi-V-firstorder}) in the case $k=2$. Substituting therein equations (\ref{KIn-t-small}b,c), (\ref{cI-0-k=l=2}), (\ref{cIm-22-fin}), we obtain
\Align$$
&m^{-4}\T_r\left[\e^{-S_\text{int}}\overline{\upstrut\bd^2\varphi(z,\bz)}_{r_0}\overline{\upstrut\d^2\varphi(z_1,\bz_1)}_0\right]|0\rangle_\nu
\notag
\\
&\qquad
=\left(-m^{-4}\ba_{-2}\a_{-2}-{\pi\tg\pi\nu\over16(1-4\nu^2)^2}b\Q
+{1\over4(1-4\nu^2)^2}\left(-\log{mr_0\over2}+\delta_\nu+{1+4\nu^2\over1-4\nu^2}-{3\over4}\right)\right)|0\rangle_\nu
\notag
\\*
&\qquad\quad
+{\nu \beta_\nu^{-2}(mr_0)^{-2-4\nu}\over2(1+2\nu)}|0\rangle_{\nu+b^2}-{\nu \beta_\nu^2(mr_0)^{-2+4\nu}\over2(1-2\nu)}|0\rangle_{\nu-b^2}
\notag
\\*
&\qquad\quad
-{(1-4\nu)\beta_\nu^{-4}(mr_0)^{-8\nu}\over32(1-2\nu)^2(1+2\nu)}|0\rangle_{\nu+2b^2}
-{(1+4\nu)\beta_\nu^4(mr_0)^{8\nu}\over32(1+2\nu)^2(1-2\nu)}|0\rangle_{\nu-2b^2},
\label{d2phi-bd2phi-vec}
$$
where $\delta_\nu$ is defined in (\ref{delta-nu-def}). Here we applied the equivalence (\ref{deltaQ-equivalence-b2}) to the divergent terms. Now let us apply the prescription (\ref{log-to-power}), (\ref{dkphi-bdkphi-vac-contrib}) to the logarithmic term and define
\Align$$
\left[\d^2\varphi\,\bd^2\varphi\,V_\nu\right]_\text{ren}
&=\overline{\upstrut\bd^2\varphi(z,\bz)}_{r_0}\overline{\upstrut\d^2\varphi(z_1,\bz_1)}_0V_\nu+{m^4(mr_0)^{4b^2}\over16b^2(1-4\nu^2)^2}V_\nu
\notag
\\*
&\quad
-{\nu \beta_\nu^{-2}m^4(mr_0)^{-2-4\nu}\over2(1+2\nu)}V_{\nu+b^2}+{\nu \beta_\nu^2m^4(mr_0)^{-2+4\nu}\over2(1-2\nu)}V_{\nu-b^2}
\notag
\\*
&\quad
+{(1-4\nu)\beta_\nu^{-4}m^4(mr_0)^{-8\nu}\over32(1-2\nu)^2(1+2\nu)}V_{\nu+2b^2}
+{(1+4\nu)\beta_\nu^4m^4(mr_0)^{8\nu}\over32(1+2\nu)^2(1-2\nu)}V_{\nu-2b^2}\biggr|_{r_0\to0}.
\label{d2phi-bd2phi-ren-def}
$$
From this definition we immediately deduce that
\eq$$
\left\langle\left[\d^2\varphi\,\bd^2\varphi\,V_\nu\right]_\text{ren}\right\rangle={b^{-2}m^4\over16(1-4\nu^2)^2}+O(b^0).
\label{d2phi-bd2phi-vac}
$$
The leading contribution confirms the exact result of \cite{Fateev:1998xb}. The contribution of order $b^0$ cannot be established in the leading order of the perturbation expansion.

Let us compare the expression (\ref{d2phi-bd2phi-vec}) with the bootstrap form factor
\Align$$
f^{c_{-2}\bc_{-2}}_\nu(\{\theta_i\}_N)
&=(\sqrt\pi b)^{-N}\sum_{\{\ve_i=\pm1\}_N}\e^{\i\pi({1\over2}-\nu)\sum_i\ve_i}
\left(-\sum^N_{i,j=1}\ve_i\ve_j\e^{2\theta_i-2\theta_j}+{1\over\sin^2\pi p}\right)
\notag
\\*
&=\left(\ctg^2\pi\nu\,\sum_{i\ne j}\e^{2\theta_i-2\theta_j}-N+{1\over\sin^2\pi p}\right)F_\nu(\{\theta_i\}_N).
\label{fc2bc2-fin}
$$
The term proportional to $\a_{-2}\ba_{-2}|0\rangle_\nu$ corresponds to the first term in the parentheses of the above expression, while the term proportional to $\Q$ corresponds to the second term. The last term corresponds (in the leading order in $b$) to the last term in (\ref{dkphi-bdkphi-vac-contrib}). Thus we have
\eq$$
V^{c_{-2}\bc_{-2}}_\nu={16(1-4\nu^2)^2\over\pi^2b^2m^4}\,\d^2\varphi\,\bd^2\varphi\,V_\nu+O(b^{-2})V_\nu.
\label{Vc2bc2-rel}
$$
Note that the first term is a sum of two contributions. The first one is of the same order as that of $V_\nu$, but its form factors depend on rapidities. The second term is proportional to $b^{-4}V_\nu$. Its form factors are rapidity independent, and the coefficient is determined by the renormalization (\ref{d2phi-bd2phi-ren-def}). This coefficient is consistent with the last term in (\ref{fc2bc2-fin}). However, a comparison of subleading coefficients cannot be carried out in the leading order of the semiclassical expansion, which is expressed by the undefined second term of (\ref{Vc2bc2-rel}).

\section{Discussion}

In this paper we considered form factors of local operators semiclassically. We found that beyond exponential operators the classical limit of form factors is ill\-/defined, since the form factors contain in the same order of $b\sim\hbar^{1/2}$ both classical and quantum contributions, including those from interaction terms in the action. The situation becomes especially complicated for operators that contain both right and left chiralities, which demands a renormalization procedure. It turns out that in some cases the contributions of \emph{lower} orders in $b$ (but of a specific fixed form) should be added for the correct renormalization. It means that our construction should be extended by performing calculations beyond the leading contributions to avoid ambiguities.

We performed calculations by decomposing quantum fields in a special basis, which is a generalization of the radial quantization of the free massless boson. This construction demands studying asymptotics and integrals of novel special functions, which generalize the Bessel functions (so to say the `sinh\-/Bessel function'). These functions can be effectively studied with the help of the Fredholm determinant solutions to the classical sinh\-/Gordon equation. We hope that this representation will be useful for calculating the contributions of higher orders in $b^2$. Besides, the very mathematical construction of `sinh\-/Bessel functions' in terms of the Fredholm determinants looks intriguing and deserves further investigation. Thus we hope to better understand the structure of the exact bootstrap formulas for form factors on this basis in future.

\section{Acknowledgments}

The authors are grateful to M.~Bershtein, A.~Litvinov and, especially, to \fbox{Ya.~Pugai} for discussions. The work of M.L.\ was supported by the Russian Science Foundation under the grant 23\--12\--00333.

\appendix

\section{Consistency of the radial quantization construction}
\label{app:consistency}

The construction of the radial quantization (\ref{chi-reddecomp})\--(\ref{bravac-def}), (\ref{f-fact})\--(\ref{Bessel-generalized}) needs to be checked in the Hamiltonian picture. In this appendix we will use the notation $\tau=\log t$, i.e.
$$
z=m^{-1}\e^{\tau+\i\xi},
\qquad
\bz=m^{-1}\e^{\tau-\i\xi},
$$
so that $\tau$ represents the Euclidean radial time. From the quadratic part of the action $\Delta S[\chi]$ defined in (\ref{shG-action-Delta}) we obtain the radial momentum
\eq$$
\Pi(\tau,\xi)=\i{\d L^\text{E}\over\d(\d_\tau\chi)}={\i\over8\pi}\d_\tau\chi(\tau,\chi),
\label{momentum-def}
$$
where $L^\text{E}$ is the Euclidean Lagrangian density, and the radial Hamiltonian
\eq$$
H(\tau)=\int^{2\pi}_0d\xi\,\lcolon\left(\Pi(\tau,\xi)\chi(\tau,\xi)+L^E\right)\rcolon
=\int^{2\pi}_0d\xi\,\lcolon\left(4\pi\Pi^2+{(\d_\xi\chi)^2\over16\pi}+{\e^{2\tau}\over16\pi}\chi^2\ch\phi_\nu(\e^\tau)\right)\rcolon.
\label{Ham-def}
$$

First, let us check the equal\-/time commutation relations using the decomposition (\ref{chi-reddecomp}). It is straightforward to check that
$$
[\Pi(\tau,\xi'),\chi(\tau,\xi)]={\i t\over2\pi}\left(u_*'u_0-u_*u_0'+\sum_{k\ne0}{1\over2k}(u_k'u_{-k}-u_ku_{-k}')\e^{\i k(\xi-\xi')}\right).
$$
From (\ref{Besgen-Wronskian}), (\ref{u-t-zero}) we deduce that
\eq$$
t(u_*'u_0-u_*u_0')=-1,
\qquad
t(u_k'u_{-k}-u_ku_{-k}')=-2k.
\label{uk-Wronskian}
$$
Hence,
\eq$$
[\Pi(\tau,\xi'),\chi(\tau,\xi)]=-{\i\over2\pi}\sum_{k\in\Z}\e^{\i k(\xi-\xi')}=-\i\delta(\xi-\xi'\bmod{2\pi})
\label{Pi-chi-commut}
$$
in consistency with the canonical quantization. In a similar way we obtain
\eq$$
[\chi(\tau,\xi'),\chi(\tau,\xi)]=[\Pi(\tau,\xi'),\Pi(\tau,\xi)]=0.
\label{chichi-PiPi-commut}
$$

Second, let us check the consistency of the definitions (\ref{ketvac-def}), (\ref{bravac-def}) of the radial vacua. The easiest way to do it is to compare the pair correlation function of the field $\chi(x)$ defined in the radial quantization scheme with that defined by the functional integral with the quadratic part of the action $\Delta S[\chi]$. For multipoint correlation functions the coincidence will follow from the Wick theorem.

To this end, note that the free fluctuation propagator calculated under such conventions is equal to
\eq$$
\langle \chi(x)\chi(x')\rangle_\text{free}=
\langle \infty|\T_r[\chi(x)\chi(x')]|0\rangle_\nu=
\Cases{
4\sum\limits_{k\in\mathbb Z}\e^{i(\xi-\xi')} K_{\nu,|k|}(mr)I_{\nu,|k|}(mr'),& r>r',\\
4\sum\limits_{k\in\mathbb Z}\e^{i(\xi-\xi')}K_{\nu,|k|}(mr')I_{\nu,|k|}(mr),\quad & r<r'.
}
\label{freeprop}
$$
Here we have used (\ref{f-fact}), (\ref{u-KIn-final}) to express the modes $f_k(x)$, $\bar f_k(x)$, $f_*(x)$ in terms of the special basis of solutions of the sinh-Bessel equation (\ref{Bessel-generalized}) introduced in (\ref{KIn-t-small-leading}).

In the functional integral formalism, such a propagator is directly related to the Green function $G\left( x,x'\right)$ of the differential operator
\eq$$
\mathsf H=-\nabla^2+m^2\ch\phi_\nu( mr),
\label{opH}
$$
determined by the quadratic part of the action (\ref{shG-action-Delta}). The Green function solves $\mathsf H_xG(x,x')=\delta(x-x')$ and satisfies the boundary conditions imposed on the field $\chi$ with respect to both of its arguments. In the case at hand, we require $\chi(x)$ to remain finite as $|x|\to 0$ and vanish as $|x|\to\infty$.   Since $\mathsf H$ commutes with the angular momentum operator $\mathsf L=-i\d_\xi$ (the generator of translations in the radial quantization framework), the action of $\mathsf H$ leaves the eigenspaces of $\mathsf L$ invariant. This leads to the decomposition
\eq$$
G(x,x')={1\over{2\pi}} \sum_{k\in\mathbb Z}G_{|k|}(mr,mr')\e^{ik(\xi-\xi')}.
\label{GFdec}
$$

The radial Green function $G_n(t,t')$ (note that by convention $n\ge0$) solves the ODE
\eq$$
\left(-{d^2\over dt^2}-{1\over t}{d\over dt}+{n^2\over t^2}+V(t)\right)G_n(t,t')={1\over t}\,\delta(t-t')
\label{radGFdef}
$$
with $V(t)=\ch\phi_\nu(t)$. It satisfies the symmetry $G_n(t,t')=G_n(t',t)$, remains regular as $t\to0$  and vanishes as $t\to\infty$. 
It then becomes clear that 
\eq$$
G_n(t,t')= C_n\Cases{
I_{V,n}(t) K_{V,n}( t'),\qquad & t<t',\\
I_{V,n}(t') K_{V,n}(t),\qquad & t>t',}
\label{radGFv2}
$$
where $I_{V,n}(t)$ is the solution of the sinh-Bessel equation (\ref{Bessel-generalized}) that remains regular as $t\to0$ and $ K_{V,n}(t)$ is the solution that decays as $t\to\infty$. The asymptotic conditions fix these solutions uniquely up to normalization prefactors, which allows us to identify them with  $I_{\nu,n}(t)$ and $K_{\nu,n}(t)$.

The constant $C_n$ has to be chosen so as to ensure the jump of the derivative of $G(t,t')$ at $t=t'$ determined by the delta function on the right of (\ref{radGFdef}):
\eq$$
-\d_t G( t,t')\Bigr|_{t=t'+0}+\d_t G(t,t')\Bigr|_{t=t'-0}={1\over t'}.
\label{derjump}
$$
This immediately implies that $C_n^{-1}=t(K_{\nu,n}( t)I'_{\nu,n}(t)-K'_{\nu,n}( t)I_{\nu,n}(t))=1$, where the last evaluation follows from the normalization of $I_{\nu,n}(t)$, $K_{\nu,n}(t)$ set by (\ref{KI-Bessel-t-small}). All of the above straightforwardly extends to an arbitrary positive radially symmetric potential $V(t)$ satisfying $V(t\to0)=o(t^{-2})$ and $V(t\to\infty)=1$. 

One thus finds that
\eq$$
G_n(t,t')=\Cases{
I_{\nu,n}(t) K_{\nu,n}( t'),\qquad & t<t',\\
I_{\nu,n}(t') K_{\nu,n}(t),\qquad & t>t'.}
\label{radGFv3}
$$
Using this expression in (\ref{GFdec}) and comparing with (\ref{freeprop}), we reproduce the expected result
\eq$$
\langle\chi(x)\chi(x')\rangle_\text{free}=8\pi G(x,x'),
\label{GFprop}
$$
confirming thereby the consistency of the definition of the vacua (\ref{ketvac-def}), (\ref{bravac-def}).

\section{Small $t$ expansions of the generalized Bessel functions}
\label{app:smallt}

Here we give several leading terms of the small $t$ expansions of the generalized Bessel functions. The expansions for $I_{\nu,k}(t)$ are completely determined by the equation (\ref{Bessel-generalized}) and the leading asymptotics (\ref{KIn-t-small-leading}). The expansions for $K_{\nu,k}(t)$ can be found in this way with the precision $O(t^k)$ only, because further terms are sensitive to admixing the second solution $I_{\nu,k}(t)$. To obtain them, the recursion relations (\ref{Phik-Psik-recursion}) must be used.

For $|\nu|<\thalf$ we have
\subeq{\label{KIn-t-small}
\Align$$
K_{\nu,k}(t)
&=2^{k-1}\,(k-1)!\,t^{-k}\bigl(1+c^{(1)}_{\nu,-k}t^{2-4\nu}+c^{(1)}_{-\nu,-k}t^{2+4\nu}
\notag
\\*
&\quad+c^{(2)}_{\nu,-k}t^{4-8\nu}+c^{(2)}_{-\nu,-k}t^{4+8\nu}+c^{(2')}_{\nu,-k}t^4+O\left(t^{6-12|\nu|}\log t\right)\bigr)
\qquad(k>0,\ k\ne2),
\tag{\ref{KIn-t-small}a}
\\
I_{\nu,k}(t)
&={t^k\over2^k\,k!}\bigl(1+c^{(1)}_{\nu,k}t^{2-4\nu}+c^{(1)}_{-\nu,k}t^{2+4\nu}
\notag
\\*
&\quad+c^{(2)}_{\nu,k}t^{4-8\nu}+c^{(2)}_{-\nu,k}t^{4+8\nu}+c^{(2')}_{\nu,k}t^4+O\left(t^{6-12|\nu|}\right)\bigr)
\qquad(k\ge0),
\tag{\ref{KIn-t-small}b}
$$}
where
\subeq{\label{cnuk-def}
\Align$$
c^{(1)}_{\nu,k}
&={\beta_\nu^{-2}\over8(1-2\nu)(k+1-2\nu)},
\tag{\ref{cnuk-def}a}
\\
c^{(2)}_{\nu,k}
&={\beta_\nu^{-4}\over2^7(1-2\nu)^3(k+1-2\nu)},
\tag{\ref{cnuk-def}b}
\\
c^{(2')}_{\nu,k}
&=-{k(k+1+4(k+3)\nu^2)\over2^6(k+2)(1-4\nu^2)^2((k+1)^2-4\nu^2)},
\tag{\ref{cnuk-def}c}
$$}
and $\beta_\nu$ is defined by (\ref{nu-B-def}). There are two exceptional cases
\Align$$
K_{\nu,0}(t)
&=-\log{t\over2}+\delta_\nu+c^{(1)}_{\nu,0}\left(-\log{t\over2}+\delta_\nu+{1\over1-2\nu}\right)t^{2-4\nu}
+c^{(1)}_{-\nu,0}\left(-\log{t\over2}+\delta_\nu+{1\over1+2\nu}\right)t^{2+4\nu}
\notag
\\*
&\quad+c^{(2)}_{\nu,0}\left(-\log{t\over2}+\delta_\nu+{1\over1-2\nu}\right)t^{4-8\nu}
+c^{(2)}_{-\nu,0}\left(-\log{t\over2}+\delta_\nu+{1\over1+2\nu}\right)t^{4+8\nu}
\notag
\\*
&\quad+{(1+12\nu^2)t^4\over128(1-4\nu^2)^3}+O\left(t^{6-12|\nu|}\log t\right),
\tag{\ref{KIn-t-small}c}
\\
K_{\nu,2}(t)
&=2t^{-2}\biggl(1+c^{(1)}_{\nu,-2}t^{2-4\nu}+c^{(1)}_{-\nu,-2}t^{2+4\nu}
+c^{(2)}_{\nu,-2}t^{4-8\nu}+c^{(2)}_{\nu,-2}t^{4+8\nu}
\notag
\\*
&\quad+{t^4\over16(1-4\nu^2)^2}\left(-\log{t\over2}+\delta_\nu+{1+4\nu^2\over1-4\nu^2}\right)+O\left(t^{6-12|\nu|}\log t\right)
\biggr),
\tag{\ref{KIn-t-small}d}
$$
where
\eq$$
\delta_\nu=2\log2+\thalf\txpsi(\thalf-\nu)+\thalf\txpsi(\thalf+\nu),
\qquad\txpsi(z)=(\log\Gamma(z))'.
\label{delta-nu-def}
$$
In fact, logarithmic contributions appear in all functions $K_{\nu,k}$ with even $k$, but in further terms.

\section{Liouville theory}
\label{app:Liouville}

In this appendix we use the notations $\phi_\nu$, $K_{\nu,k}(t)$ etc.\ differently from the main body of the paper, using them in the context of the Liouville model. These functions and constants are much simpler than in the case of the sinh\-/Gordon model.

Let us consider the analog of our construction in the Liouville theory
\eq$$
S^\text{L}_b[\varphi]={1\over8\pi}\int d^2x\,\left({(\d_\mu\varphi)^2\over2}+{m^2\over2b^2}(\e^{b\varphi}-1)\right).
\label{Liouv-action}
$$
Again, let us start from the classical radial Liouville equation
\eq$$
t^{-1}\d_t(t\,\d_t\phi)=\thalf\e^\phi.
\label{Liouv-radial}
$$
Its exact solution $\phi(t)=\phi_\nu(t)$ is explicit in elementary functions:
\eq$$
\e^{-\phi_\nu(t)/2}=\beta_\nu t^{2\nu}+\beta_{1-\nu}t^{2-2\nu}.
\label{phi-Liouv-exact}
$$
The coefficients $\beta_\nu$ are given by the same formula (\ref{nu-B-def}).%
\footnote{In fact, the expression (\ref{phi-Liouv-exact}) defines a solution for any function $\beta_\nu$ subject to the condition $16(2\nu-1)^2\beta_\nu \beta_{1-\nu}=-1$. The actual function $\beta_\nu$ is extracted from the requirement that the general solution in the form $\e^{-\phi(x)/2}=\beta_\nu U_1(z)U_1(\bz)+\beta_{1-\nu}U_2(z)U_2(\bz)$ is single\-/valued on the Euclidean plane provided that $U_1$, $U_2$ are two linearly independent solutions to the equation $(\d^2-T(z))U(z)=0$ with $T(z)=\sum_a\nu_a(1-\nu_a)(z-z_a)^{-2}$ such that e.g.\ $U_1(z)\simeq(z-z_1)^{\nu_1}$, $U_2(z)\simeq(z-z_1)^{1-\nu_1}$ as $z\to z_1$.}%
Note that for $0\le\nu<\thalf$ this expression exactly coincides with the small\-/distance asymptotics (\ref{phi-t-small}) of $\phi_\nu(t)$.

Similarly, consider fluctuations about the classical solution:
\eq$$
\varphi(x)=b^{-1}\phi_\nu(mr)+\chi(x),
\label{chi-Liouv-def}
$$
and define the regularized action
\Align$$
S^{\text{L},\reg}_\nu
&=\left.{1\over8\pi}\int_{\ve<mr<T}d^2x\,\left(-{(\d_\mu\phi)^2\over2}+{m^2\over2}(\e^\phi-1-\phi\e^\phi)\right)-2\nu^2\log\epsilon\,+2(1-\nu)^2\log T
\right|_{\substack{\phi(x)=\phi_\nu(mr)\\\ve\to0\\T\to\infty}}
\notag
\\
&=\left.{1\over4}\int^T_\ve dt\,t\left(-{\phi^{\prime\,2}_\nu(t)\over2}+{1\over2}(\e^{\phi_\nu(t)}-1-\phi_\nu(t)\e^{\phi_\nu(t)})
\right)-2\nu^2\log\epsilon\,+2(1-\nu)^2\log T
\right|_{\substack{\ve\to0\\T\to\infty}}
\label{Sreg-nu-Liouv-def}
$$
for $\nu<{1\over2}$. Then
\eq$$
S^\text{L}_b[\varphi]-b^{-1}\nu\varphi(0)=b^{-2}S^{\text{L},\reg}_\nu+\Delta S^\text{L}[\chi]+\const_1\cdot\nu^2+\const_2\cdot(1-\nu)^2,
\label{Liouv-action-decomp}
$$
where
\Align$$
\Delta S^\text{L}[\chi]
&={1\over8\pi}\int d^2x\left({(\d_\mu\chi)^2\over2}+{m^2\over2b^2}(\e^{b\chi}-b\chi-1)\e^{\phi_\nu(mr)}\right)
\notag
\\
&={1\over8\pi}\int d^2x\left({(\d_\mu\chi)^2\over2}+{m^2\over4}\chi^2\e^{\phi_\nu(mr)}+\cdots\right).
\label{Liouv-action-Delta}
$$
Then the considerations (\ref{chi-reddecomp})--(\ref{u-t-zero}) can be repeated verbatim. The functions $u_k(t)$ now satisfy the equation
\eq$$
u''+t^{-1}u'-\left(\thalf\e^{\phi_\nu(t)}+n^2t^{-2}\right)u=0
\label{Bessel-Liouv-generalized}
$$
with $n=k$ for $k\ne*$ and $n=0$ for $k=*$. This kind of a generalized Bessel equation admits an exact solution:
\eq$$
u_k(t)=t^{-k}{2k-2-t\phi'_\nu(t)\over2k-2+4\nu},
\qquad
u_*(t)={1\over4}{\d\phi_\nu(t)\over\d\nu}.
\label{uk-Liouv-exact}
$$
The pairs $u_k$, $u_{-k}$ provide two independent solutions for $k>0$, while $u_0$, $u_*$ provide a pair of independent solutions for $k=0$. The corresponding generalized Bessel functions are
\eq$$
K_{\nu,k}(t)=2^{k-1}\,(k-1)!\,u_k(t)\quad(k>0),
\qquad
K_{\nu,0}(t)=u_*(t),
\qquad
I_{\nu,k}(t)={1\over2^k\,k!}u_{-k}(t)\quad(k\ge0).
\label{Kn-phi-rel-Liouv}
$$
In the region $\nu<\thalf$ the small\-/distance asymptotics of this solution is $u_k(t)=t^{-k}\left(1+O\left(t^{2-4\nu}\right)\right)$, $u_*(t)=-\left(1+O\left(t^{4-4\nu}\right)\right)\log t$. The large\-/distance asymptotics are
\eq$$
u_k(t)=t^{-k}\left(K^\infty_{\nu,k}+O\left(t^{4\nu-2}\right)\right),
\quad
u_*(t)=-\left(K^\infty_{\nu,0}+O\left(t^{4\nu-2}\right)\right)\log t,
\quad
K^\infty_{\nu,k}={k+1-2\nu\over k-1+2\nu}
\quad(\nu<\thalf).
\label{uk-tlarge-Liouv}
$$
Correspondingly, $I^\infty_{\nu,k}=K^\infty_{\nu,-k}$. Evidently, the connection coefficients $K^\infty_{\nu,\pm k}$ here do not coincide with the connection coefficients defined in (\ref{KI-t-large}) for the `sinh\-/Bessel' functions (and are not even approximations thereof). In a sense, they describe renormalization of the operators $\a_{-k}$ under the influence of the exponential operator $V_\nu$ in the Liouville theory.


\providecommand{\href}[2]{#2}\begingroup\raggedright\endgroup

\end{document}